\documentclass[11pt]{article}
\usepackage[utf8]{inputenc}
\usepackage{graphicx}
\usepackage{amsmath}
\usepackage{url}
\usepackage{comment}
\usepackage[margin=1.25in]{geometry}
\usepackage[multiple]{footmisc}
\usepackage{appendix}
\usepackage[sort]{natbib}
\usepackage{todonotes}
\usepackage{soul}
\usepackage{tablefootnote}
\usepackage{placeins}
\usepackage{lscape}
\usepackage[hidelinks, hyperfootnotes=false]{hyperref}
\usepackage{bbm}
\usepackage{relsize}

\usepackage{longtable}
\usepackage{array}
\newcolumntype{C}[1]{>{\centering\arraybackslash}p{#1}}

\makeatletter
\renewcommand\hyper@natlinkbreak[2]{#1}
\makeatother

\usepackage[tableposition=below]{caption} 
\title{The Disparate Impacts of College Admissions Policies on Asian American Applicants}
\author{
    Joshua Grossman\\
    Stanford University
    \and
    Sabina Tomkins\\
    University of Michigan
    \and
    Lindsay Page\\
    Brown University
    \and
    Sharad Goel\\
    Harvard University
}

\newcommand\numapps{685,709}
\newcommand\numstudents{292,795}

\newcommand\mainagap{28\%}
\newcommand\mainsagap{49\%}
\newcommand\maineagap{17\%}

\newcommand\fullsagap{30\%}

\newcommand\precision{97\%}
\newcommand\recall{91\%}

\newcommand\pa{36\%}

\newcommand\psa{34\%}
\newcommand\pea{51\%}
\newcommand\psea{15\%}

\newcommand\pacceptw{12\%}

\newcommand\pacceptsa{10\%}
\newcommand\pacceptea{16\%}
\newcommand\pacceptsea{8\%}

\newcommand\pacceptwthirtyfour{16\%}
\newcommand\pacceptsathirtyfour{9\%}
\newcommand\pacceptthirtyfourgap{43\%}

\newcommand\poverthirtytwo{93\%}

\newcommand\appendixappcoverage{\ref{fig:n_apps_by_year}}
\newcommand\appendixreliable{\ref{fig:summary-stats-reliable-transcript}}
\newcommand\appendixsummarystats{\ref{fig:summary-stats-by-race}}
\newcommand\appendixobserved{\ref{fig:observed-unobserved-vars}}
\newcommand\appendixcovariates{\ref{fig:basket-year-model-covariates}–\ref{fig:location-model-covariates}}
\newcommand\appendixrobustness{\ref{fig:robustness-east-asian}–\ref{fig:robustness-southeast-asian}}

\newcommand\appendixnocalifornia{\ref{fig:p_admit_by_state}}
\newcommand\appendixstateswithmoreasian{\ref{fig:p_asian_high_scoring_by_state}}
\newcommand\appendixhighschool{\ref{fig:p_admit_by_high_school}}
\newcommand\appendixhistoricalenrollment{\ref{fig:p_race_enrolled_by_year}}
\newcommand\appendixlegacybyrace{\ref{fig:p_legacy_by_race}}

\date{}
\begin{document}

\maketitle
\thispagestyle{empty}

\begin{abstract}
    There is debate over whether Asian American students are admitted to selective colleges and universities at lower rates than white students with similar academic qualifications.
    However, there have been few empirical investigations of
    this issue,
    in large part due to a dearth of data.
    Here we present the results from analyzing  
    \numapps~applications from Asian American and white students
    to a subset of selective U.S. institutions
    over five application cycles, beginning with the 2015--2016 cycle.
    The dataset does not include admissions decisions, and so we construct a proxy based in part on enrollment choices.
    Based on this proxy, we estimate the odds that Asian American applicants 
    were admitted to at least one of the schools we consider
    were \mainagap~lower
    than the odds for white students with similar test scores, grade-point averages, and extracurricular activities.
    The gap was particularly pronounced for students of South Asian descent (\mainsagap~lower odds).
    We trace this pattern in part to two factors.
    First, many selective
    colleges openly give preference to the children of alumni,
    and we find that white applicants were substantially more likely to have such legacy status than Asian applicants, especially South Asian applicants.
    Second, after adjusting for observed student characteristics, 
    the institutions we consider appear less likely to admit students from geographic regions with relatively high shares of applicants who are Asian.
    We hope these results inform ongoing discussions on the equity of college admissions policies.
\end{abstract}

\section*{Introduction}

Over the last several decades, questions have been raised over whether selective colleges in the U.S.\ discriminate against Asian American applicants in admissions decisions~\citep{chun1992civil, takagi1992retreat, espenshade2004admission, espenshade2009no, long2004race, park2019interest, 2019students, arcidiacono2022asian, gelman2019statistics}.
In the 1980s, Brown and Stanford
formed committees to audit their own admissions policies and practices~\citep{chun1992civil,takagi1992retreat}.
Brown found evidence of discrimination in its admissions process; Stanford did not find clear evidence of bias, but could not fully explain its lower acceptance rates of Asian American applicants relative to white students.
A 1990 report by the U.S.\ Department of Education's Office of Civil Rights (OCR) investigated allegations that Harvard capped the number of Asian American students it admitted~\citep{chun1992civil}.
OCR found no evidence of an Asian quota, but concluded that Asian American applicants were less likely to be admitted than white students with similar academic qualifications.
OCR further found that this disparity largely disappeared once recruited athletes and the children of alumni (``legacies'') were excluded from its analysis, suggesting the gap in acceptance rates was driven by Harvard's stated preference for admitting students from these two groups~\citep{hurwitz2011impact, park2019interest, chetty2023legacy}.
Most recently, in a 2023 decision,
the Supreme Court ruled that
Harvard engaged in unconstitutional racial balancing, holding the Asian American share of admitted students to approximately 20\%---though
Harvard denied doing so.
In the more than 30 years since the OCR investigation, there have been limited third-party, applicant-level empirical analyses of potential discrimination in college admissions decisions against Asian American applicants.
Over this time span, both the demographics of the United States and the educational landscape have changed substantially.
Asian American representation among K–12 public school students has more than doubled, increasing from 3\% in 1993 to 7\% in 2020~\citep{nowicki2022k}, and the overall admission rate to Harvard has dropped from 18\% in 1990 to 5\% in 2020~\citep{1990-admit-rate, 2020-admit-rate}.
These changes suggest a need to reexamine college admissions policies for potential disparate impacts on Asian American applicants.

Here we analyze \numapps~first-year college applications submitted by \numstudents~Asian American and white students to 
a subset of U.S. institutions with relatively low admit rates and relatively high yield rates.
All of the applications we consider were submitted via a national postsecondary application platform 
over five application cycles, from the 2015--2016 cycle to the 2019--2020 cycle.\footnote{%
    Each of the institutions
    we consider receives the majority of first-year applications from students applying via the 
    application platform (Table \appendixappcoverage). 
}
We exclude students who attend a high school outside of the United States or who report primary citizenship outside of the United States.
Given the complex patterns of immigration and marked heterogeneity in experiences across subgroups,
we disaggregate our analysis 
by three regions of origin self-reported by the Asian American applicants in our dataset: South Asia, East Asia, and Southeast Asia.\footnote{%
Once applicants indicate being ``Asian'', they have the option to select one or more of 9 countries of origin; they can additionally indicate being ``Other East Asian'', ``Other South Asian,'' or ``Other Southeast Asian''.
We classify the listed countries as follows: China, Japan, and Korea as East Asia; India and Pakistan as South Asia; and the Philippines, Vietnam, Cambodia, and Malaysia as Southeast Asia. 
3\% of Asian applicants select countries that span multiple regions. 
In these cases, we randomly assign one of the spanned regions.
2\% of Asian applicants do not select a country of origin. 
These students are excluded from the analysis.
}
To preserve confidentiality, we
focus on broader patterns rather than on individual institutions, and we report aggregate results across the combined set of colleges and universities we consider.
In particular, our main outcome of interest is whether applicants were admitted to at least one of these institutions.
One limitation of our analysis is that we do not directly observe admissions decisions, and so we infer these decisions based on enrollment choices, as described below.

After excluding students who we infer to be recruited athletes,  we estimate that South Asian applicants  
had \mainsagap~lower odds of admission
to the subset of schools we consider
than white applicants with comparable test scores, high school grade-point averages, and extracurricular activities.
We estimate that both East Asian and Southeast Asian applicants had \maineagap~lower odds of admission to these schools.
After additionally adjusting for whether a student applied early to 
any considered college or university, the student's high school, and whether the student is a legacy applicant, 
we estimate that Southeast Asian students were accepted at similar rates to white students, and that East Asian students had 10\% lower odds of admission than white students. 
But, we estimate that South Asian applicants 
still had \fullsagap~lower odds of acceptance to these institutions
than white students
after adjusting for all available information in our data.
We note, however, that we do not have access to all materials submitted by and about applicants, such as essays, letters of recommendation, alumni interviews, and admission officer ratings.
Finally, we explore how the relative share of Asian American and white enrollees might change at 
the colleges and universities we consider
under various hypothetical admission policies.
Under a policy that admits students solely on the basis of standardized test scores and participation in extracurricular activities---and holding fixed the combined number of enrolled Asian American and white students---we estimate that enrollment of South Asian students and East Asian students
would increase substantially, while the number of Southeast Asian students would remain approximately the same. 

Concerns about the disparate impacts of college admissions policies on Asian American students are often entangled with discussions about affirmative action~\citep{west2016obscuring, parkinequality, hughes2016causation, takagi1992retreat, kim2022applying, karabel2005chosen, gersen2017uncomfortable, antonovics2013affirmative, gelman2019statistics}.
At their core, however, these two issues---affirmative action and 
differences in the admission rates of similarly qualified white and Asian American students---are conceptually distinct.
In particular, during the time period we consider, institutions could have admitted Asian American applicants at rates comparable to similarly qualified white students 
while still giving preference to applicants from groups underrepresented in higher education.\footnote{%
As of 2023,
explicit racial preferences in college admissions are no longer legally permissible~\citep{supreme2023sffa}. 
}

\section*{Data description}

Our analysis is based on applications submitted through 
a national postsecondary application platform.
The data we use contain detailed, anonymized information on each student, including 
race and gender;
standardized test scores (ACT and/or SAT);
high school grade-point average (GPA);
Advanced Placement (AP) exam scores;
structured descriptions of their extracurricular activities (e.g., the number of hours they spent participating in various clubs or sports);
the location and other characteristics of the high school they attended;
whether their parents attended college, and, if so, the colleges they attended;
whether they received an application fee waiver 
(a proxy for financial need);
the set of 
colleges to which they applied via the platform;
and whether they applied early action or early decision to any of the institutions we consider (Table~\appendixobserved).
If a student took the SAT, we convert their SAT score to an equivalent ACT score to facilitate comparisons between applicants and aid interpretation.\footnote{%
If students took both the ACT and SAT, or took either test more than once, we choose the highest ACT-equivalent score achieved on a single test.
}
Although we have quite detailed individual-level data, we do not have access to the full set of application materials, including student essays, letters of recommendation, or intended major.
We also do not have access to internal college evaluations, such as interviewer ratings.

We approximate admissions decisions by first inferring enrollment decisions. 
We infer enrollment by observing the school to which a high school counselor sent a student's official high school transcript, 
information that is collected by the platform.
(NB: official transcripts typically are required by colleges to formalize acceptance decisions.)
We then infer that 
students were admitted to at least one of the schools we consider if, and only if, they sent a transcript to (i.e., ultimately enrolled in) one of those schools.
This inference rests on an assumption that students who were admitted to at least one of the schools we consider
ultimately attended 
one of those schools.
While imperfect, three points suggest this process yields results that are suitably accurate for our purposes. 
First, we assessed the quality of our enrollment inference by matching 5,000 randomly selected 
applicants to the schools we consider to be their true enrollments as reported by the National Student Clearinghouse.
We find that the estimated precision of our enrollment inference strategy is \precision~with an estimated recall of \recall. 
We further find that accuracy is comparable across race groups (see the \nameref{methods} section in the Appendix).
Second, the schools we consider have relatively high yield rates, suggesting that admission to these schools is strongly correlated with enrollment.
Finally, we find qualitatively similar results with an estimation strategy that holds under the weaker assumption that enrollment is independent of race, conditional on acceptance and other observed student characteristics (see the \nameref{rrappendix} section in the Appendix for details).

Our study pool is comprised of \numapps~applications
 submitted by \numstudents~students to the colleges and universities we consider in the 2015--2016 through the 2019--2020 application cycles.
We include Asian and white applicants who attended a U.S.\ high school,
excluding students from high schools for which we cannot reliably infer college enrollment (see the \nameref{methods} section in the Appendix and Table~\appendixreliable). 
We cannot identify 
athletic recruits with certainty, but we exclude from our sample students who appear to be athletic recruits based on the timing of their applications and their reported extracurricular activities (see the \nameref{methods} section in the Appendix). 
Within our study pool, \pa~of applicants self-identify as Asian, with \pea, \psea, and \psa~of these students self-identifying as East Asian, Southeast Asian, and South Asian, respectively.
Finally, we supplement our 
data from the platform with public high school data from the Common Core of Data (CCD), private high school data from the Private School Universe Survey (PSS), and rurality data at the ZIP code level from the Economic Research Service of the U.S.\ Department of Agriculture. 

\section*{Results}
\begin{figure}[t]
    \centering
    \includegraphics[width=4in]{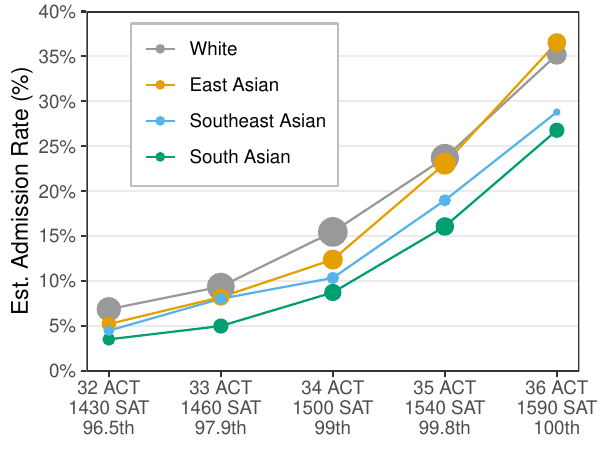}
    \caption{
        \emph{Estimated rate of admission to at least one of the selective
        institutions we consider as a function of standardized test score, for Asian American applicants and white applicants in the study pool.
        Asian American applicants typically were admitted at lower rates than white applicants with identical test scores, with the largest gap for South Asian students.
        Among 
        admits in our study pool who report ACT or SAT scores, \poverthirtytwo~have ACT (or ACT-equivalent) scores at or above 32. 
        Percentiles are derived from all students who took the ACT in 2018~\citep{act-2018}. 
        Point sizes are proportional to the number of applicants in each group.}
    }
    \label{fig:p_accept_by_region}
\end{figure}

Among applicants to the colleges and universities we consider,
we estimate that \pacceptea~of East Asian, \pacceptsea~of Southeast Asian, and \pacceptsa~of South Asian students were admitted to at least one 
of these institutions,
compared to \pacceptw~of white applicants.
While these aggregate admissions rates differ by race and ethnicity, they do not account for differences in qualifications across groups.
For example, Asian American applicants had, on average, higher standardized test scores than white applicants (Table \appendixsummarystats). 
As a first step to account for these differences, in Figure~\ref{fig:p_accept_by_region} we show estimated admissions rates by standardized test score for Asian American applicants and white applicants.
We find that Asian American students were
admitted at consistently lower rates than white applicants with comparable test scores, 
with the largest gap for South Asian applicants.
For instance, among applicants with an 
ACT (or ACT-equivalent) score of 34---placing them in the 99th percentile of test takers---we estimate that \pacceptwthirtyfour~of white students were admitted compared to \pacceptsathirtyfour~of South Asian students, a relative gap of \pacceptthirtyfourgap.

Standardized test scores are one factor among many that colleges consider when determining whom to admit. 
Additional criteria that we are able to observe include high school grade-point average (GPA), participation in extracurricular activities,
legacy status, and the state in which each applicant's high school is located.
To understand the extent to which these other considerations may explain the observed disparities in admissions rates, we fit a series of nested logistic regression models of the following form:

\begin{align*}
    & \Pr (Y_i=1) = \text{logit}^{-1}(
    \beta_0\! +\! 
    \beta_S \mathbbm{1}_S\! +\!
    \beta_E \mathbbm{1}_E\! +\!
    \beta_{SE} \mathbbm{1}_{SE}\! +\!
    X_i \beta_X ),
\end{align*}

\noindent where $Y_i$ is a binary variable indicating whether applicant $i$ was admitted to any 
college or university we consider;
$\mathbbm{1}_S$, $\mathbbm{1}_E$, and $\mathbbm{1}_{SE}$ indicate 
whether the applicant identified as South Asian, East Asian, or Southeast Asian, respectively; and $X_i$ is a vector of additional covariates (e.g., test scores and GPA) that we vary across models, with $\beta_X$ the corresponding vector of coefficients.
Our key coefficients of interest are
$\beta_S$, $\beta_E$, and $\beta_{SE}$, 
which yield estimates of the gap in admissions rates between white applicants and Asian American applicants in the three Asian subgroups that we consider.
We find similar results if we fit separate models comparing white applicants to applicants in each Asian subgroup individually (Tables~\appendixrobustness). 

Table~\ref{fig:main-regression-table} shows, for nine models that include different subsets of control variables, the fitted coefficients for each of the three Asian subgroups (see also Tables~\appendixcovariates).
Coefficients are exponentiated for ease of interpretation as odds ratios.
The first model includes only fixed effects for the application season and the subset of 
colleges (or application ``basket'') to which the student applied---among the full set of colleges we consider---facilitating comparisons among groups of students who applied in the same year and to the same subset of colleges. 
The corresponding coefficients are thus akin to raw 
admissions odds ratios
across groups, without adjusting for differences in applicant credentials. 

The second and third models in Table~\ref{fig:main-regression-table} additionally adjust for measures of academic preparation, 
including SAT/ACT alone (Model 2) and, additionally, GPA, AP test scores, and SAT II subject test scores (Model 3).
These academic-preparation models corroborate the visual pattern in Figure~\ref{fig:p_accept_by_region}:
we estimate that Asian American students---especially South Asian students---had substantially lower odds of admission
than white students with similar test scores and related academic credentials.
These disparities largely persist when we progressively adjust for extracurricular activities (Model 4);
gender and family characteristics, like whether the student received an application fee waiver (Model 5);
and whether the student applied early (Model 6). 

    \begin{table*} \centering 
    \scriptsize
    \vspace{-6em}
    \centerline{
        \begin{tabular}{@{\extracolsep{0pt}}lccccccccc} 
    \\[-1.8ex]\hline 
    \hline \\[-1.8ex] 
     & \multicolumn{9}{c}{Outcome: Inferred acceptance to at least one college or university we consider} \\ 
    \cline{2-10} 
    \\[-1.8ex] & \multicolumn{9}{c}{} \\ 
     & Basket+Year & SAT/ACT & GPA+AP+SAT2 & Activities & Sex+Family & Early App & Legacy & Location+HS & All \\ 
    \\[-1.8ex] & (1) & (2) & (3) & (4) & (5) & (6) & (7) & (8) & (9)\\ 
    \hline \\[-1.8ex] 
     South Asian & 0.66$^{***}$ & 0.56$^{***}$ & 0.59$^{***}$ & 0.51$^{***}$ & 0.51$^{***}$ & 0.52$^{***}$ & 0.61$^{***}$ & 0.60$^{***}$ & 0.70$^{***}$ \\ 
      & (0.01) & (0.01) & (0.01) & (0.01) & (0.01) & (0.01) & (0.01) & (0.01) & (0.02) \\ 
      & & & & & & & & & \\ 
     Southeast Asian & 0.64$^{***}$ & 0.73$^{***}$ & 0.78$^{***}$ & 0.83$^{***}$ & 0.81$^{***}$ & 0.84$^{***}$ & 0.88$^{***}$ & 0.94 & 1.02 \\ 
      & (0.02) & (0.02) & (0.02) & (0.03) & (0.03) & (0.03) & (0.03) & (0.03) & (0.04) \\ 
      & & & & & & & & & \\ 
     East Asian & 1.11$^{***}$ & 0.86$^{***}$ & 0.85$^{***}$ & 0.83$^{***}$ & 0.79$^{***}$ & 0.73$^{***}$ & 0.90$^{***}$ & 0.88$^{***}$ & 0.90$^{***}$ \\ 
      & (0.02) & (0.01) & (0.01) & (0.02) & (0.01) & (0.01) & (0.02) & (0.02) & (0.02) \\ 
      & & & & & & & & & \\ 
      \hline \\[-1.8ex] 
     Aggregated Asian & 0.88$^{***}$ & 0.74$^{***}$ & 0.75$^{***}$ & 0.72$^{***}$ & 0.69$^{***}$ & 0.67$^{***}$ & 0.79$^{***}$ & 0.79$^{***}$ & 0.85$^{***}$ \\ 
     (separate model) & (0.01) & (0.01) & (0.01) & (0.01) & (0.01) & (0.01) & (0.01) & (0.01) & (0.02) \\  
    \hline \\[-1.8ex] 
    Basket+Year & X & X & X & X & X & X & X & X & X \\ 
    \hline \\[-1.8ex]
    SAT/ACT &   & X & X & X & X & X & X & X & X \\ 
    \hline \\[-1.8ex]
    GPA+AP+SAT2 &   &   & X & X & X & X & X & X & X \\ 
    \hline \\[-1.8ex]
    Activities &   &   &   & X & X & X & X & X & X \\ 
    \hline \\[-1.8ex]
    Sex+Family &   &   &   &   & X & X & X & X & X \\ 
    \hline \\[-1.8ex]
    Early App &   &   &   &   &   & X &   &   & X \\ 
    \hline \\[-1.8ex]
    Legacy &   &   &   &   &   &   & X &   & X \\ 
    \hline \\[-1.8ex]
    Location+HS &   &   &   &   &   &   &   & X & X \\ 
    \hline \\[-1.8ex]
    Observations & 292,795 & 292,795 & 292,795 & 292,795 & 292,795 & 292,795 & 292,795 & 292,795 & 292,795 \\ 
    In-sample AUC & 0.65 & 0.75 & 0.79 & 0.81 & 0.82 & 0.85 & 0.83 & 0.84 & 0.88 \\ 
    White base rate & 12\% & 12\% & 12\% & 12\% & 12\% & 12\% & 12\% & 12\% & 12\% \\ 
    \hline \\[-1.8ex] 
    \multicolumn{10}{r}{$^{*}$p$<$0.05; $^{**}$p$<$0.01; $^{***}$p$<$0.001} \\ 
    \end{tabular}} 
        \caption{
                \emph{Estimated conditional odds of admission to at least one college or university we consider for Asian American applicants in the study pool relative to white applicants.
                Coefficients are estimated via logistic regression, and are exponentiated for ease of interpretation as odds ratios.
                After adjusting for test scores and extracurricular activities (Model 4), South Asian students had 48\% lower estimated odds of admission relative to white students,
                with East Asian and Southeast Asian applicants exhibiting smaller but statistically significant gaps (17\% lower odds of admission).
                These disparities appear to be explained in part by legacy preferences (Model 7) and geography (Model 8).
                The ``Aggregated Asian'' coefficients are computed from separate models that do not separate students into Asian subgroups.
        }}
        \label{fig:main-regression-table}
    \end{table*} 

\begin{figure}[t]
    \centering
    \includegraphics[width=4in]{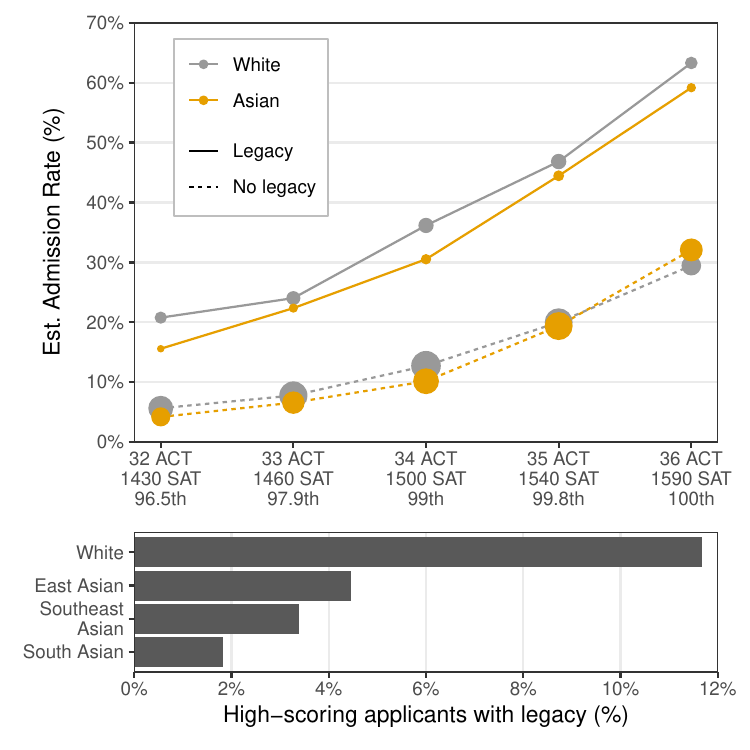}
    \caption{
        \emph{Estimated rate of admission to at least one college or university we consider for white applicants and Asian American applicants with high ACT or SAT scores.
        Across test scores, we estimate that applicants with a parent who attended one of the selective institutions we consider as an undergraduate are more than twice as likely to be admitted than non-legacy applicants with the same test scores.
        The bottom panel shows the proportion of applicants with high test scores who have legacy status, disaggregated by race.
        High-scoring white applicants are three to six times more likely to have legacy status than high-scoring Asian American applicants, suggesting
        white applicants disproportionately benefit from a boost in admission rates afforded to those with legacy status.
    }}
    \label{fig:p_accept_by_legacy}
\end{figure}

Next, with Model 7, we account for whether a student is the child of an alum. 
After adjusting for legacy status---in addition to all of the above mentioned factors---we see large reductions in the estimated disparities in acceptance rates for all three Asian subgroups we consider.
Figure~\ref{fig:p_accept_by_legacy}
helps explain this result.
The top panel of the figure
shows 
estimated admission rates for Asian American applicants and white applicants conditional on legacy status and test scores.\footnote{%
In Figure~\ref{fig:p_accept_by_legacy},
we follow convention and define legacy status to mean an applicant had at least one parent who attended one of the colleges or universities we consider
as an undergraduate, and the student applied to the institution(s) that their parent(s) attended.
In our regression models, we additionally adjust for other familial connections to the included colleges and universities,
like a parent attending graduate school there or having two parents with undergraduate degrees from the same school
}\footnote{%
In prior work examining the effect of legacy status on admission to elite institutions, \citet{hurwitz2011impact} found that the magnitude of the legacy effect is larger in models that account for the application components that we do not observe.
}
For a given test score, we estimate that applicants---both white and Asian American---with legacy status were more than twice as likely to gain admission than applicants without legacy status. 
In the bottom panel of Figure~\ref{fig:p_accept_by_legacy}, we present prevalence of legacy status among applicants with an ACT-equivalent test score of 32 or above, mirroring the focus of the upper panel. 
Here, we observe that white applicants were approximately three times more likely to have legacy status than East Asian and Southeast Asian applicants, and almost six times more likely than South Asian students. 
Thus, even though estimated acceptance rates conditional on test score and legacy status were similar across race and ethnicity, white students appear to benefit from being substantially more likely to have legacy status.

In theory, the higher estimated admissions rates that we observe for legacy applicants may stem both from admissions practices that favor the children of alumni and from the potentially greater social capital of legacy students.
We note, however, that 
Model 5 adjusts for whether an applicant 
had a parent who attended a top-50 institution  (based on 2019 U.S.\ News rankings) not included in the subset of colleges on which we focus,
or attended one of the colleges in our subset to which the student did not apply---proxies for having high social capital distinct from legacy status specifically.
The change in disparities that we observe moving from Model 5 to Model 7 thus appears attributable to the specific benefits of having legacy status, rather than the more generalized benefits of high social capital.

\begin{figure}[t]
    \centering
\includegraphics[width=4in]{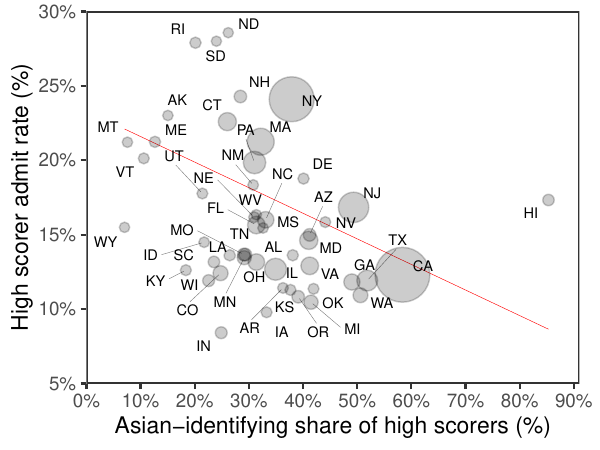}
    \caption{
        \emph{For each U.S. state, overall estimated admission rate to at least one institution among the subset of selective schools we consider for white applicants and Asian applicants with an ACT-equivalent score at or above 32, with the proportion of high-scoring white and Asian applicants who identify as Asian on the horizontal axis.
        Point sizes are proportional to the number of high-scoring white and Asian applicants from the state who applied to one of the institutions we consider.
         The red least-squares regression line is weighted by the same count of applicants.
        States with a greater share of Asian American applicants have, on average, lower estimated admission rates for high-scoring applicants.
    }}
    \label{fig:p_accept_by_state}
\end{figure}

Finally, we examine the relationship between estimated acceptance rates and geography.
For each state, Figure~\ref{fig:p_accept_by_state} displays the estimated admission rate of high-achieving applicants---with ACT-equivalent scores of 32 or above---to the fraction of applicants from that state who were Asian American.
In computing this proportion, we limit to white applicants and Asian American applicants,
and point sizes are proportional to the total number of high-scoring white and Asian American applicants in each state.
The negatively sloped regression line shows that states with a larger fraction of Asian American applicants tended to have lower estimated admission rates.
Further, states with a higher proportion of Asian American applicants tended to have higher average test scores, suggesting the geographic trend is not driven by a gap in academic achievement (Figure~\appendixstateswithmoreasian).
This geographic pattern also persists when we
exclude applicants from California, and when we disaggregate the data to the level of high school instead of state (Figures \appendixnocalifornia~and~\appendixhighschool).

Model 8 in Table~\ref{fig:main-regression-table}---which adjusts for location as well as academic and extracurricular performance but not legacy status---shows that these apparent geographic preferences account for much of the admissions gap between white and Asian American applicants. 
Model 9, the last one we consider, adjusts for all application information available to us, including both legacy status and geography.
After adjusting for this rich set of covariates, we see that the estimated admissions gap between Southeast Asian and white applicants largely disappears, though we still find that white students have higher estimated odds of admission than otherwise similar East Asian and South Asian applicants.
It is unclear what may account for these remaining disparities, though it bears repeating that admissions officers have access to more complete application materials than do we, 
including letters of recommendation, essays, and interview assessments.

We conclude our analysis by exploring how the relative share of Asian American students at the institutions we consider might change under various hypothetical admissions policies.
In line with our analysis above, we restrict our attention to white students and Asian American students. 
Specifically, we hold fixed the combined number of students in these groups (approximately mirroring historical admissions outcomes, as shown in Figure~\appendixhistoricalenrollment), and so any increases in Asian American enrollment necessarily imply decreases in enrollment of white students.
Any exercise of this sort is inherently speculative---in part because changes in admissions policies could alter application behavior---but we still believe it is informative to gauge the approximate magnitude of effects.

\begin{figure}[t]
    \centering
\includegraphics[width=4in]{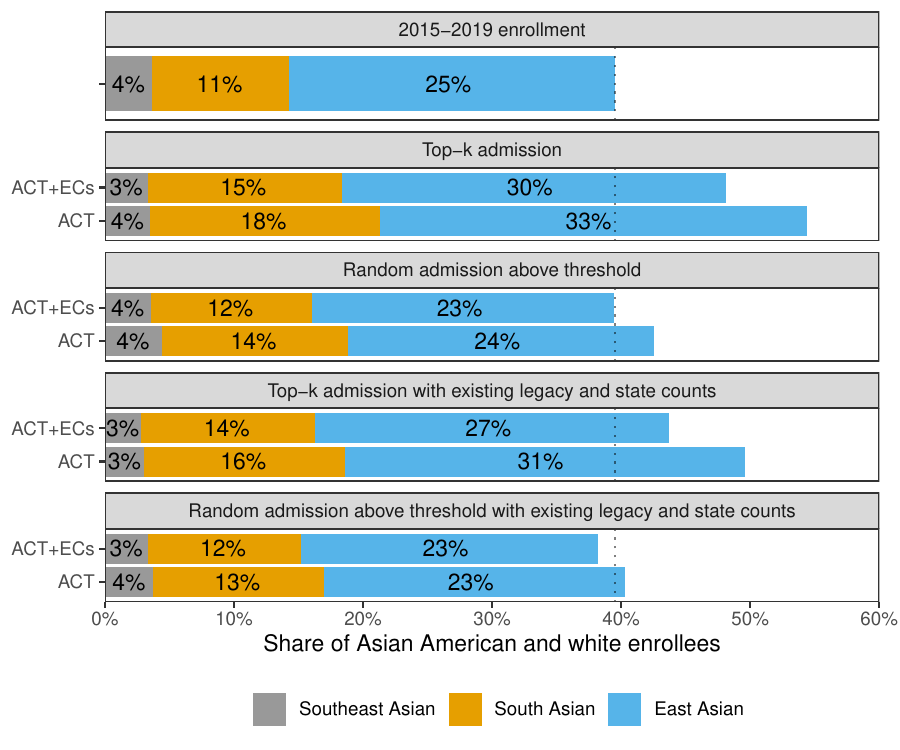}
    \caption{
        \emph{Estimated enrollment of Asian American students 
        at the institutions we consider under eight hypothetical admissions policies,
        with the top panel showing the actually observed demographic composition in our historical data.
        In all cases, we consider only the subset of Asian American students and white students, and so increases in Asian American enrollment correspond to decreases in the enrollment of white students.
        In most instances, the hypothetical policies we consider lead to an increase in enrollment of Asian American students, including those that preserve the number of legacy students and the number of enrollees from each state in the historical data. 
    }}
    \label{fig:policies}
\end{figure}

As a baseline, the top row of Figure~\ref{fig:policies} shows the estimated share of enrollees in our data from the three Asian subgroups 
of interest.
The rest of the figure shows the estimated share of enrollees from these subgroups under eight hypothetical admissions policies that are divided into four categories.
In the first category---which we call ``top-$k$'' policies---we imagine admitting  students with the highest ACT-equivalent scores, with ties broken randomly. 
In the second category, ``random above threshold,'' we consider policies that randomly admit students above an ACT-equivalent score $t$ such that admitted students have a mean score equal to that of actual enrollees~\citep{sandel2020tyranny}. 
For both of these categories we consider two variants: 
the ``ACT'' variant selects from the entire applicant pool of the schools we consider, while the ``ACT+ECs'' variant selects only from applicants with 
at least as many hours of reported extracurricular (EC) activities over four years of high school as the median of the hours reported by all enrollees. 
Under all four policies, we estimate the same or
larger shares of Asian American students
compared to what we observe in the data.
Asian American students report, on average, fewer extracurricular hours than white applicants, so the ACT+ECs policy variant admits fewer Asian American applicants than the ACT variant.

The final two categories we consider
investigate outcomes under hypothetical policies that maintain both the current number of enrollees from each state and the total number of enrollees with legacy.
Specifically, we first divide our historical data into 102 (2 x 51) cells consisting of legacy and non-legacy applicants from each U.S. state and Washington, D.C.;
we then in turn apply each of the four policies described above to each of the 102 cells, 
ensuring for each cell that the number of students enrolled under the hypothetical policies matches the historical enrollment numbers.
With these added legacy and geographic constraints, the share of Asian American enrollees is smaller than under the unconstrained analogs, as expected given our results above.
But, even with these constraints,
the number of Asian American enrollees across policies is still similar to or larger than the status quo.

\section*{Discussion}

Based on a large-scale analysis of applications to a subset of selective U.S. colleges and universities, 
it appears that that Asian American students 
were less likely to be admitted than white students with comparable academic credentials and extracurricular activities, a
disparity that is particularly pronounced for South Asian students. 
It further appears that much---though not all---of this gap is attributable to admissions practices that favor the children of alumni and apparent geographic preferences. 
These disparities likely stem from a complex set of objectives that universities work to balance, and are not necessarily driven by explicit or implicit racial preferences.
Nonetheless, our results prompt questions about the equitable design of college admissions policies.

In our primary analysis, we excluded applicants who we inferred were recruited athletes, under the assumption that filling sports teams is a hard constraint for many universities, and that
doing so involves qualitatively distinct admissions criteria.
We note, though, that athletic recruits are disproportionately likely to be white rather than Asian American: in our study pool, white applicants outnumber Asian American applicants by a factor of about two to one, but among inferred recruits, white applicants outnumber Asian American applicants by a factor of four to one. 
As a result, if we do not proactively exclude recruited athletes from our analysis,
we find an even larger gap in the estimated admissions rates between Asian American students and white students with comparable academic credentials (Tables \appendixrobustness).

Our results are subject to two key limitations.
First, we have imperfect information on college admissions decisions. 
In our analysis, we infer admissions decisions from enrollment choices, where we assume that students who applied to but did not ultimately attend one of the selective schools we consider were not admitted to any of those schools.
This assumption only allows us to approximately reconstruct admissions decisions.
However, 
given the relatively high yield rates of the universities we consider,
we believe this assumption is suitably accurate for our analysis.
Further, we find qualitatively similar results under an alternative estimation strategy that rests on the weaker assumption that enrollment decisions are independent of race, conditional on acceptance and other observed student characteristics (see the \nameref{rrappendix} section in the Appendix).
Finally, our results remain largely the same if we eliminate any one school from our analysis (Tables \appendixrobustness), suggesting the robustness of our results to the exact subset of schools we consider.

Second, we do not have access to each student's complete application materials. Specifically, we do not observe a student's intended major, essays, teacher recommendations, transcripts, interview ratings, and admission officer ratings.
It is thus possible that students who we observe to have similar academic and extracurricular credentials are in fact different in important ways that are revealed in these other materials.
We note, however, that results made public through litigation 
suggest that---at least in the case of Harvard's admissions practices---the disparities we identify persist after adjusting for several additional markers of academic and extracurricular excellence, including admission officer ratings of each applicant's academics, extracurriculars, teacher recommendations, and counselor recommendations
(cf. Figure 6.1 in~\citep{arcidiacono-expert-report}, Model 4 to Model 5, Table B.7.1 and B.7.2 show coefficients).\footnote{%
Expert testimony provided in the Harvard case indicates that disparities in admission rates at Harvard are reduced after adjusting for admission officers'
assessments of an applicant's ``personal qualities'' and admission officers'
``overall rating'' of an applicant.
There is worry, however, that assessments of ``personal qualities''
are more subjective than ratings of academic and extracurricular achievements, are less clearly connected to merit, and may be influenced by implicit or explicit racial biases. 
Further, ``overall ratings'' are so closely tied to the final admissions decision,
that we would expect adjusting for them would mask any disparities~\citep{jung2018omitted}. 
}
Further, \citet{kim2022applying} finds that Asian American and white college applicants with similar academic credentials receive letters of recommendation that are ``broadly similar in content and tone.'' 

Discussions of college admissions practices impacting Asian Americans often revolve around affirmative action. 
But, as we noted at the start, these issues are conceptually distinct.
In theory, one can both implement affirmative action policies that 
maintain the share of students on campus from groups that 
are underrepresented in higher education 
while simultaneously admitting Asian American students at the same rate as white students with similar academic and extracurricular credentials. 
In such a case, we would expect the number of enrolled white students to decrease, not the number of racial minorities. 
During the time period we examined, affirmative action was widely used for shaping the diversity of college campuses, meaning the scenario described above was an option available to college administrators.
Thus, at the very least, our results shed light on past admissions choices and their consequences for Asian American college applicants.
Now that affirmative action is legally prohibited, institutions will need to reconsider how applicants are evaluated in order to ensure equitable admissions processes and to maintain diverse campuses.
For example, existing decision-making processes that afford preference to the children of alumni appear to not only disadvantage Asian Americans but also other racial minorities (Figure~\appendixlegacybyrace).
Looking ahead, we hope our findings facilitate ongoing discussions about the design and implementation of equitable admissions policies.

\newpage

\bibliographystyle{plainnat}
\bibliography{bib}

\clearpage
\appendix
\section{Appendix}
\renewcommand\thefigure{\thesection\arabic{figure}}  
\renewcommand\thetable{\thesection\arabic{table}}    
\setcounter{figure}{0} 
\setcounter{table}{0} 

\nameref{methods} 
\begin{itemize}
    \item The first subsection describes the data filters used to construct the study pool.
    \item The second subsection describes how a record of a sent transcript is deemed as a reliable or unreliable signal of enrollment.
    \item The third subsection describes our validation of sent transcripts as a signal of enrollment using true records of enrollment from the National Student Clearinghouse.
    \item The fourth subsection describes how we attempt to identify potential athletic recruits among the applicant pool.
\end{itemize}

\noindent ``Estimating Admission Rates'' summarizes a complementary analysis contingent on the weaker assumption that enrollment is independent of race, conditional on acceptance and other observed student characteristics. \\

\noindent Table~\ref{fig:n_apps_by_year} shows the proportion of publicly reported applications reported by the institutions in the main analysis that were submitted via the national postsecondary application platform. \\

\noindent Table~\ref{fig:summary-stats-reliable-transcript} shows summary statistics for applicants from high schools with reliable and unreliable records of sent transcripts.  \\

\noindent Table~\ref{fig:summary-stats-by-race} shows summary statistics for white, East Asian, South Asian, and Southeast Asian applicants in the study pool. \\

\noindent Table~\ref{fig:observed-unobserved-vars} lists relevant covariates observed and unobserved by the authors and by the application platform.  \\

\noindent Tables~\ref{fig:basket-year-model-covariates} through~\ref{fig:location-model-covariates} list and describe the covariates used in the main model specifications. \\

\noindent Tables~\ref{fig:robustness-east-asian} through~\ref{fig:robustness-southeast-asian} describe robustness checks of the main model specification. \\

\noindent Tables~\ref{fig:replication_data_white} through~\ref{fig:aggregated_replication_data} show the number of applicants in various coarsened groups defined by race and ethnicity, test score, legacy status, region of the country, and enrollment signal. \\

\begin{figure}[t]
    \centering
    \includegraphics[scale=1]{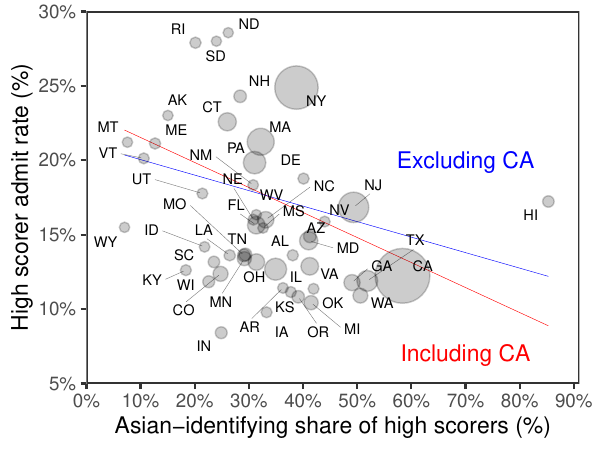}
    \caption{
        For each U.S. state, estimated rate of admission to the selective schools we consider for white applicants and Asian applicants with an ACT-equivalent score at or above 32, with the proportion of high-scoring white and Asian applicants who identify as Asian on the horizontal axis.
        Point sizes are proportional to the number of high-scoring white applicants and Asian applicants to the considered schools whose high school is located in the given state.
        The red least-squares regression line is weighted by the same count of high-scoring white and Asian American applicants from each state.
        The blue line excludes applicants from California.
        States with a greater share of Asian American applicants have, on average, lower admission rates for high-scoring applicants. 
        This pattern holds even if applicants from California are excluded.
    }
    \label{fig:p_admit_by_state}
\end{figure}

\begin{figure}[t]
    \centering
    \includegraphics[scale=1]{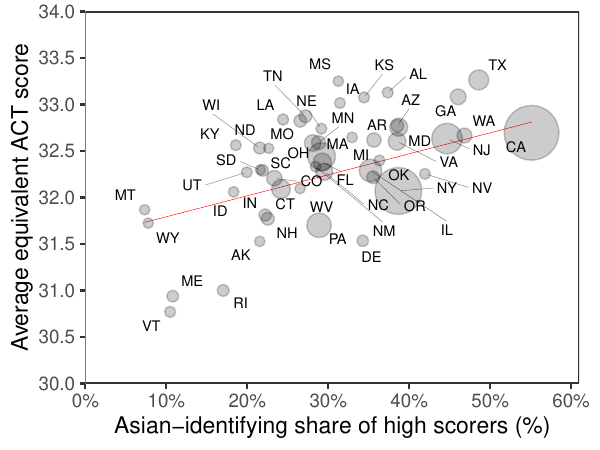}
    \caption{
        For each U.S. state, mean equivalent ACT score among applicants who reported an ACT score, with the proportion of high-scoring white and Asian applicants who identify as Asian on the horizontal axis.
        Hawaii is excluded from the plot due to its exceptionally high share of Asian American applicants.
        Hawaii's mean equivalent ACT score is 31.6.
        Point sizes are proportional to the number of high-scoring white applicants and Asian applicants to the considered schools whose high school is located in the given state.
        The red least-squares regression line is weighted by the same count of white and Asian American applicants from each state.
    }
    \label{fig:p_asian_high_scoring_by_state}
\end{figure}

\begin{figure}[t]
    \centering
    \includegraphics[scale=1]{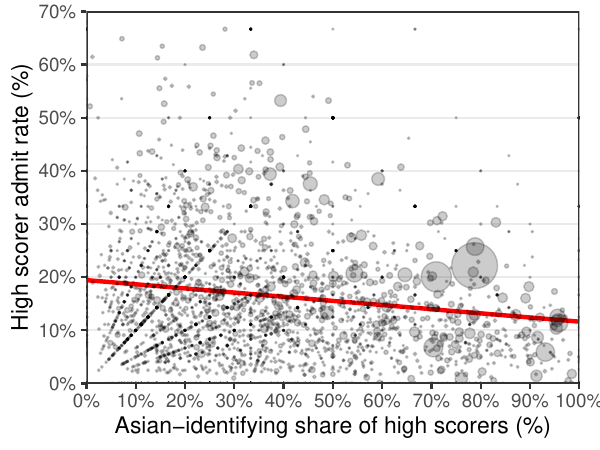}
    \caption{
        For each high school in the study pool, rate of admission to any of the selective schools we consider for white applicants and Asian American applicants with an ACT-equivalent score at or above 32, with the proportion of high-scoring white and Asian American applicants who identify as Asian American on the horizontal axis.
        Point sizes are proportional to the number of high-scoring white applicants and Asian applicants to the considered institutions who attend the given high school.
        The red least-squares regression line is weighted by the same count of high-scoring white and Asian American applicants from the given high school.
        High schools with a greater share of Asian American applicants have, on average, lower admission rates for high-scoring applicants. 
    }
    \label{fig:p_admit_by_high_school}
\end{figure}

\noindent Figures~\ref{fig:p_admit_by_state} through~\ref{fig:p_admit_by_high_school} show robustness checks of the state-level analysis in the main text. \\

\begin{figure}[t]
    \centering
    \includegraphics[scale=.75]{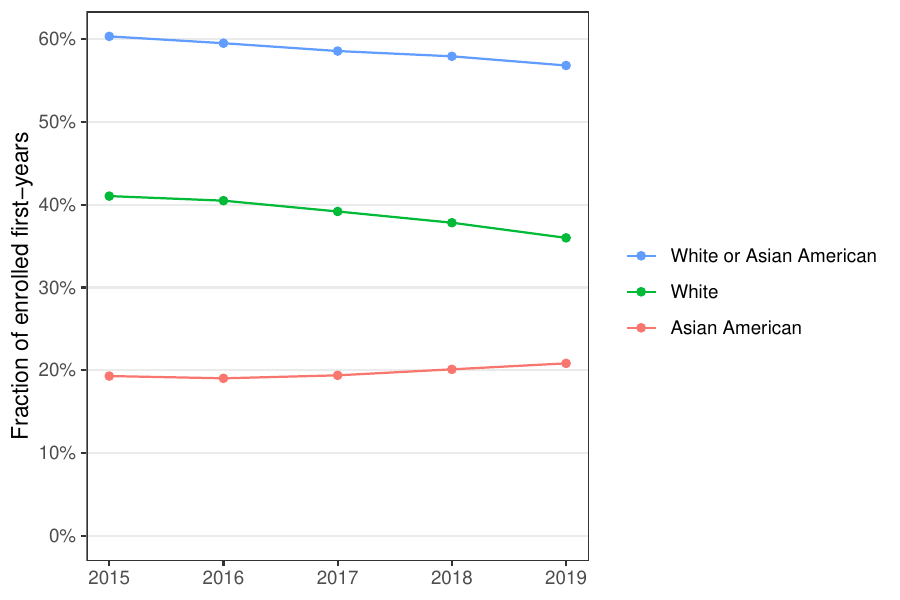}
    \caption{
        Fraction of observed enrollments to the selective schools we consider attributed to Asian American and white enrollees. 
        The overall share of enrollments attributed to Asian American and white enrollees decreases slightly over the five years included in the analysis.
        The hypothetical policies described in the main analysis assume that this share remains approximately constant regardless of application year.
    }
    \label{fig:p_race_enrolled_by_year}
\end{figure}

\begin{figure}[t]
    \centering
    \includegraphics[scale=.75]{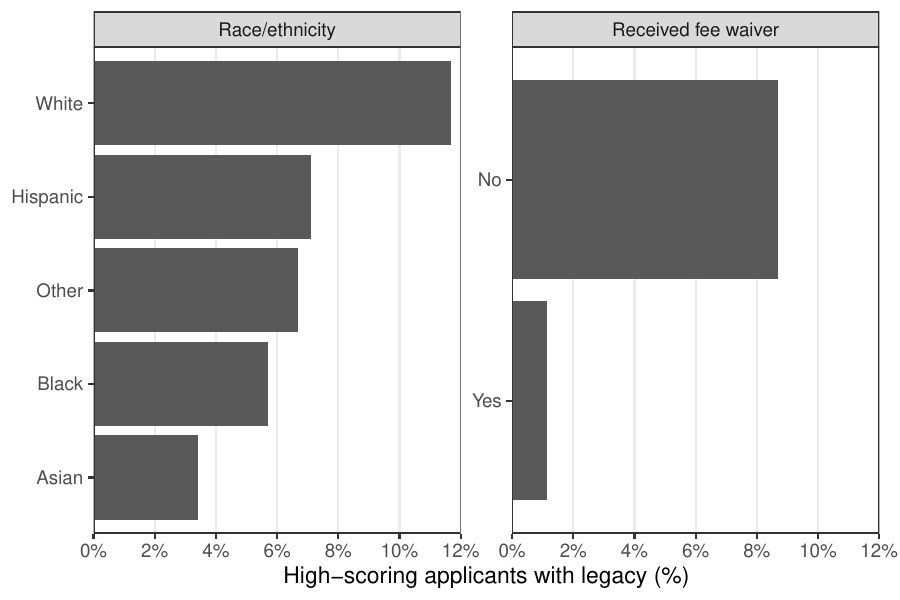}
    \caption{
        The proportion of applicants with one or more parents who attended one of the selective schools we consider as an undergraduate, by race/ethnicity and fee waiver status.
        The pool of applicants in this figure is the same as the main analysis, but does not apply the filters for race or ethnicity.
        White applicants are the most likely to have legacy, and Asian American applicants are the least likely. 
    }
    \label{fig:p_legacy_by_race}
\end{figure}

\noindent Figures~\ref{fig:p_race_enrolled_by_year} and~\ref{fig:p_legacy_by_race} support the hypothetical policy results in the main text. \\

\newpage

\section*{Methods}
\label{methods}

\subsection*{Data filtering}

We begin our core analysis with the 551,292 South Asian, East Asian, Southeast Asian, and white students who submitted at least one application to one of the selective schools we consider via the national postsecondary application platform in the 2015–2016 application cycle through the 2019–2020 cycle.
We then filter to the 449,564 applicants who attended a high school in a U.S.\ state or the District of Columbia, and who did not report citizenship outside of the United States.
We next limit to the 297,417 applicants who attended high schools with official transcript sends that, to the best of our knowledge, accurately reflect an intention to enroll.
Finally, for our main analysis, we restrict to the 292,795 applicants who, to the best of our knowledge, are not athletic recruits.

\subsection*{Identifying high schools with reliable transcript sending behavior}

Typically, when an applicant intends to enroll in a particular college to which they were admitted, their high school must submit an official transcript to the college.
Many high schools use the same portal to submit official transcripts, and the platform observes when an applicant's transcript is sent via this portal.
While the platform does not observe acceptances or enrollments, high school transcript sends serve as a highly accurate enrollment proxy for the subset of applicants who meet the following conditions:

\begin{itemize}
    \item First, the applicant's high school must use the transcript sending platform.
    In other words, we only include applicants whose high school sent at least one transcript via the platform in the same year the applicant applied.
    \item Second, transcript sends must be targeted to specific colleges.
    If a high school counselor does not track the intended enrollment of a particular applicant, they may indiscriminately sent final transcripts to every college to which the applicant submitted an application.
\end{itemize}

We define a high school's transcript sending behavior as ``reliable'' in a given year if the high school submits the same number of transcripts as applications for fewer than 5\% of applicants who submit at least two applications.
In this definition, we do not consider applicants who submit one application and one transcript, as we cannot reliably guess whether the student intended to enroll or the transcript was sent indiscriminately by their counselor. 
In Tables \ref{fig:robustness-east-asian}-\ref{fig:robustness-southeast-asian}, we replicate the main results with thresholds other than 5\%, finding qualitatively similar results.

Among the applicants from high schools who exhibit reliable transcript sending behavior, we further exclude the applicants who submitted the same number of transcripts as applications and submitted at least two applications, since these students' counselors likely sent the transcripts indiscriminately. 

\subsection*{Verifying transcript send enrollment signal with NSC data}

Using a stratified random sample of 5,000 enrollments obtained from the National Student Clearinghouse (NSC), we find that our enrollment heuristic has nearly perfect precision (97\%) and high, but not perfect, recall (91\%).
The 5,000 sampled applicants were selected from the study pool, which includes only those applicants whose high school met the threshold for reliable transcript sending behavior in the given application year.

Among the first stratum of 2,500 applicants with a transcript sent to at least one of the selective schools we consider, we were able to match 2,336 (93\%) to NSC enrollments.
We find that 2,271 actually enrolled in one of the selective schools we consider within a year of admission.
Thus, the estimated precision of the enrollment proxy is 97\%.
Precision is nearly identical across race groups: white students have a precision of 97\%, while Asian American students have a precision of 97.1\%.
Precision is also similar across application years, high school states, and application fee waiver status. 

We are ultimately interested in the likelihood of admission to our specific set of selective schools.
Thus, the transcript sending signal is arguably superior to knowledge of actual enrollment, as an admitted student with the intention to enroll may later decide not to enroll. 
If we were to use the student's true enrollment as a signal of admission to one of the selective schools, we would falsely assume that the student was not admitted to one of the selective schools.
Thus, the true precision of our enrollment signal may be even higher than 97\%. 

For the second stratum of 2,500 applicants who applied to one of the selective institutions we consider but did not send a transcript to one of these institutions, we were able to match 2,294 (92\%) to NSC enrollments.
43 (2\%) of the 2,294 matched applicants ended up enrolling at one of the selective institutions we consider within a year of applying.
We attribute this discrepancy to a number of potential factors: 
students may be admitted off the waitlist at one of the selective schools we consider; individual counselors may not use the transcript sending portal even if the rest of the high school uses the portal; or students may transfer to one of the selective schools we consider after initially being rejected. 
To determine the source of the discrepancy, we disaggregate by whether the matched applicant sent a transcript to any school on the platform.

1,247 of the 2,294 matched applicants sent a transcript to a school on the platform outside of the specific set of selective schools we consider.
Of these 1,247 applicants, 13 actually ended up enrolling at one of these selective schools within a year of applying, but 10 of these 13 enrolled first at the school to which they sent a transcript.
We assume that these 10 students transferred to one of the selective schools we consider after initially being rejected, so we exclude them from the error calculation, as our proxy for rejection is assumed to be correct for the application year.
The remaining three applicants may have been admitted off the waitlist at one of the schools we consider after initially committing to a different school.
In the study pool, 139,888 applicants sent a transcript to a school on the platform outside of the specific set of colleges and universities we consider.
Thus, from these applicants, we estimate 139,888*(3/1,247) = 337 unobserved enrollments at the considered schools.

The remaining 1,057 of the 2,294 matched applicants did not send a transcript to any school on the platform.
30 (3\%) of the 1,057 applicants ended up at one of the selective schools we consider.
We attribute these 30 enrollments to the idiosyncratic counselor behavior described above.
In the study pool, 117,138 applicants did not send a transcript to any school on the platform.
Thus, we estimate 117,138*(30/1,057) = 3,325 unobserved enrollments from these applicants.

In sum, in the study pool, we observe 35,769 enrollments to the selective schools we consider. 
We estimate 337 + 3,325 = 3,662 unobserved enrollments.
Thus, our estimated recall is 35,769/(35,769+3,662) = 91\%.
Given that our estimated recall is based on discrepant enrollments of only 33 matched applicants, we cannot meaningfully evaluate recall across groups. 

\subsection*{Identifying potential athletic recruits}

In order to field competitive athletic teams, universities often recruit students with exceptional athletic ability.
University admission offices may have a hard constraint of filling athletic teams with a sufficient number of talented student athletes.
Admissions decisions for student athletes are primarily the choice of athletic coaches, who are incentivized to offer admission to recruits with the greatest athletic ability who meet or exceed the minimum academic qualifications for admission.\footnote{%
See ``Before Recruiting in Ivy League, Applying Some Math'' [New York Times, 2011] and ``A Little Secret: Athletics at the Most Selective Colleges and Universities in the Nation: College and University Admissions, Part III'' [Huffington Post, 2011].
}
Further, student athletes are typically admitted early.\footnote{%
See ``Varsity athletes, admissions and enrollment at top colleges'' [Washington Post, 2019]
}
We thus attempt to exclude students who, to our knowledge, may be athletic recruits, as the admission process for student athletes differs considerably from that of typical applicants.
As a robustness check, we repeat the main analysis without excluding potential recruits, finding qualitatively similar results (Tables \ref{fig:robustness-east-asian}-\ref{fig:robustness-southeast-asian}).

While we do not observe true athletic recruitment status, we have access to detailed information about each applicant's extracurricular participation.
We know not only the extracurricular activities of each applicant, but also the number of years participated in each activity during high school, the order in which those activities are reported in their application, and whether the applicant intends to continue participation in the activity in college.
We assume that student athletes will list their athletic participation as the first activity in their application.
We further assume that student athletes will have participated in their first-listed sport during all four years of high school, and that they intend to continue participating in their first-listed sport in college.
Finally, 
we assume that student athletes will only apply to one college or university in an early round, and that they will always send a transcript to that one institution.

Among the 40,391 white and Asian American applicants who submitted a transcript to one of the institutions we consider in the main analysis and who also attended a high school in the U.S. with reliable transcript-sending behavior, 4,622 applicants were potential recruits (11\%).
While we cannot formally verify that these students were actually recruited by the schools we consider, 11\% is in line with public estimates of the fraction of selective university enrollees who are student athletes.\footnote{%
As above, see ``Varsity athletes, admissions and enrollment at top colleges'' [Washington Post, 2019]
}

\FloatBarrier

\newpage

\section*{Estimating Admission Rates}
\label{rrappendix}
In our main analysis, we assume that students who were admitted to one of the selective institutions we consider ultimately attended one of those institutions. In this way, we could infer admissions decisions from enrollment choices---which we can in turn accurately impute by looking at the institution to which a student sent their final high school transcript.
Here we describe an alternative estimation strategy that holds under the weaker assumption that enrollment choices are independent of race conditional on acceptance and other observable student characteristics.

Denote by $A$ the event that a particular applicant is admitted to one of the schools we consider, where $A=1$ if the applicant is admitted and $A=0$ if the applicant is not admitted. 
Denote by $E$ the analogous enrollment event. 
Finally, denote by $R$ the race of the applicant, and by $W$ a set of non-race covariates.
Now, suppose we are interested in comparing the admission probability of an applicant with race $R$ and another applicant of race $R'$ with identical non-race covariates $W$. 
We can express this comparison as a risk ratio:
\begin{align*}
    \frac{\Pr (A=1 \mid W, R')}{\Pr (A=1 \mid W, R)}.
\end{align*}
Without observing admission outcomes, the above ratio cannot be estimated directly.
But, suppose we assume that $E \perp \! \! \! \perp R \mid A=1,W$. 
In other words, conditional on acceptance and all observed non-race covariates, the decision to enroll in one of the considered institutions is independent of race. 
Then,

\begin{align*}
    \Pr (E=1 \mid W,R) & = \Pr (A=1 \mid W,R) \cdot \Pr(E=1 \mid A=1,W,R) \\
    & = \Pr (A=1 \mid W,R) \cdot \Pr(E=1 \mid A=1,W),
\end{align*}
and so
\begin{align*}
    \Pr (A=1 \mid W,R) & = \frac{\Pr (E=1 \mid W,R)}{\Pr(E=1 \mid A=1,W)}.
\end{align*}

Applying this result to the acceptance risk ratio:

\begin{align*}
    \frac{\Pr (A=1 \mid W, R')}{\Pr (A=1 \mid W, R)} & = \frac{\Pr (E=1 \mid W, R')}{\Pr (E=1 \mid A=1, W)} \cdot \frac{\Pr (E=1 \mid A=1, W)}{\Pr (E=1 \mid W, R)} \\
    & = \frac{\Pr (E=1 \mid W, R')}{\Pr (E=1 \mid W, R)}.
\end{align*}

\noindent
Thus, by assuming that enrollment is independent of race conditional on acceptance and non-race covariates, we can estimate the acceptance ratio using only data on enrollment.
Averaging over $W$, we have
\begin{align}
\label{eq:rr}
    \mathlarger{\mathlarger{\sum}}_{W} \frac{\Pr (A=1 \mid W, R')}{\Pr (A=1 \mid W, R)} \cdot \Pr(W)
    = \mathlarger{\mathlarger{\sum}}_{W} \frac{\Pr (E=1 \mid W, R')}{\Pr (E=1 \mid W, R)} \cdot \Pr(W).
\end{align}

Importantly, the right-hand side of Eq.~\eqref{eq:rr} can be estimated directly from enrollment choices, as done with the main models in our analysis (Table~\ref{fig:main-regression-table}).
In particular, take $R$ to be white students and $R'$ to be, in turn, the three Asian subgroups we consider.
Then, after adjusting for test scores, GPA, and extracurricular activities (i.e., by using Model 4 in the main text), we estimate that the average acceptance ratio is 0.58 for South Asian applicants, 0.85 for East Asian applicants, and 
0.89 for Southeast Asian applicants.
These estimates align with the results we report in Table~\ref{fig:main-regression-table}, corroborating our main analysis.

\FloatBarrier

\newpage

\begin{longtable}{|c|c|c|}
  \hline
 & Academic year & Proportion \\ 
  \hline
1 & 2015-2016 & 99\% \\ 
  2 & 2016-2017 & 98\% \\ 
  3 & 2017-2018 & 96\% \\ 
  4 & 2018-2019 & 93\% \\ 
  5 & 2019-2020 & 92\% \\ 
   \hline
\caption{
    Approximate proportion of all publicly reported applications to the selective schools we consider that were submitted via the application platform, by academic season. 
    The share of applications submitted via the platform has decreased in recent years as alternative platforms have become more popular. 
}
\label{fig:n_apps_by_year}
\end{longtable}

\FloatBarrier

\newpage

\begin{longtable}{|r|c|c|c|}
  \hline
Variable & All & Included & Excluded \\ 
  \hline
Tot. applicants & 444,420 & 292,795 & 151,625 \\ 
  Prop. sent transcript & 10\% & 12\% & 7\% \\ 
  Prop. white & 64\% & 64\% & 66\% \\ 
  Prop. Asian American & 36\% & 36\% & 34\% \\ 
  Prop. East Asian & 17\% & 18\% & 15\% \\ 
  Prop. South Asian & 12\% & 12\% & 12\% \\ 
  Prop. Southeast Asian & 6\% & 6\% & 8\% \\ 
  Mean num. apps submitted anywhere & 7.9 & 8.4 & 7.1 \\ 
  Mean num. apps submitted to subset & 2.3 & 2.3 & 2.2 \\ 
  Prop. applied early & 39\% & 42\% & 34\% \\ 
  Prop. w/ legacy & 6\% & 7\% & 3\% \\ 
  Mean ACT score & 32.1 & 32.4 & 31.4 \\ 
  Prop. unreported ACT & 14\% & 14\% & 13\% \\ 
  Mean standardized GPA & 0.8 & 0.8 & 0.8 \\ 
  Prop. unreported GPA & 14\% & 16\% & 10\% \\ 
  Mean num. AP tests & 4 & 4.1 & 3.8 \\ 
  Median activity hours & 3196 & 3236 & 3107 \\ 
  Median sports hours & 480 & 540 & 400 \\ 
  Prop. female & 53\% & 53\% & 54\% \\ 
  Prop. first generation & 14\% & 12\% & 19\% \\ 
  Prop. using fee waiver & 15\% & 12\% & 20\% \\ 
  Prop. rural HS & 6\% & 4\% & 9\% \\ 
  Prop. private HS & 22\% & 27\% & 13\% \\ 
  Median grad. class size & 333 & 319 & 366 \\ 
  Prop. from California & 17\% & 16\% & 19\% \\ 
  Prop. from Texas & 5\% & 4\% & 6\% \\ 
  Prop. from Florida & 3\% & 3\% & 4\% \\ 
  Prop. from New York & 13\% & 15\% & 9\% \\ 
   \hline
\caption{
    Summary statistics for the `Included' applicants who attend high schools with reliable transcript-sending behavior, the `Excluded' applicants who do not, and the combined set of `All' applicants.
    On average, the `Included' applicants submit more applications, apply early with a greater likelihood, are more likely to have legacy status, have higher standardized test scores, have more extracurricular hours, are more likely to play sports, are less likely to use application fee waivers, are more likely to attend urban and private high schools, and have smaller graduating class sizes.
    We re-run the main regression by inversely weighting the probability that a given applicant attends a high school with reliable transcript behavior, finding qualitatively similar results (Tables \ref{fig:robustness-east-asian}-\ref{fig:robustness-southeast-asian}, `Reweighted' model variant). 
} 
\label{fig:summary-stats-reliable-transcript}
\end{longtable}

\FloatBarrier

\newpage

\begin{longtable}{|r|c|c|c|c|c|c|}
  \hline
Variable & All & White & Asian & E Asian & S Asian & SE Asian \\ 
  \hline
Tot. applicants & 292,795 & 186,079 & 106,716 & 53,856 & 36,389 & 16,471 \\ 
  Prop. sent transcript & 12\% & 12\% & 13\% & 16\% & 10\% & 8\% \\ 
  Prop. white & 64\% & 100\% & 0\% & 0\% & 0\% & 0\% \\ 
  Prop. Asian American & 36\% & 0\% & 100\% & 100\% & 100\% & 100\% \\ 
  Prop. East Asian & 18\% & 0\% & 50\% & 100\% & 0\% & 0\% \\ 
  Prop. South Asian & 12\% & 0\% & 34\% & 0\% & 100\% & 0\% \\ 
  Prop. Southeast Asian & 6\% & 0\% & 15\% & 0\% & 0\% & 100\% \\ 
  Mean num. apps & & & & & & \\ 
  submitted anywhere & 8.4 & 8.1 & 9 & 9 & 9.5 & 7.7 \\ 
Mean num. apps & & & & & & \\ 
  submitted to subset & 2.3 & 2.1 & 2.8 & 3 & 2.9 & 2.3 \\ 
  Prop. applied early & 42\% & 41\% & 44\% & 50\% & 42\% & 33\% \\ 
  Prop. w/ legacy & 7\% & 10\% & 3\% & 4\% & 2\% & 3\% \\ 
  Mean ACT score & 32.4 & 32.2 & 32.9 & 33.3 & 32.8 & 31.4 \\ 
  Prop. unreported ACT & 14\% & 17\% & 10\% & 9\% & 11\% & 11\% \\ 
  Mean standardized GPA & 0.8 & 0.8 & 0.8 & 0.8 & 0.7 & 0.8 \\ 
  Prop. unreported GPA & 16\% & 16\% & 15\% & 16\% & 15\% & 14\% \\ 
  Mean num. AP tests & 4.1 & 3.6 & 5 & 5.3 & 5.1 & 4 \\ 
  Median activity hours & 3236 & 3384 & 2975 & 3131.7 & 2862 & 2688 \\ 
  Median sports hours & 540 & 728 & 240 & 318 & 162 & 240 \\ 
  Prop. female & 53\% & 52\% & 53\% & 54\% & 51\% & 56\% \\ 
  Prop. first generation & 12\% & 9\% & 16\% & 18\% & 10\% & 25\% \\ 
  Prop. using fee waiver & 12\% & 8\% & 19\% & 19\% & 14\% & 30\% \\ 
  Prop. rural HS & 4\% & 5\% & 1\% & 2\% & 1\% & 2\% \\ 
  Prop. private HS & 27\% & 31\% & 20\% & 20\% & 17\% & 23\% \\ 
  Median grad. class size & 319 & 280 & 400 & 400 & 403 & 372 \\ 
  Prop. from California & 16\% & 11\% & 24\% & 26\% & 16\% & 31\% \\ 
  Prop. from Texas & 4\% & 3\% & 5\% & 4\% & 7\% & 5\% \\ 
  Prop. from Florida & 3\% & 3\% & 2\% & 1\% & 3\% & 3\% \\ 
  Prop. from New York & 15\% & 14\% & 15\% & 17\% & 13\% & 12\% \\ 
   \hline
\caption{
    Summary statistics for the race and ethnicity groups included in the analysis.
    White applicants are more likely to have legacy status than Asian applicants, have a greater number of extracurricular hours, on average, and are more likely to attend smaller and private high schools. 
    East and South Asian applicants have, on average, higher standardized test scores and take more AP tests than white and Southeast Asian applicants. 
} 
\label{fig:summary-stats-by-race}
\end{longtable}

\FloatBarrier

\newpage

\newgeometry{left=0.5in, right=0.5in}

\begin{longtable}{|l|p{6.5cm}|l|l|}
  \hline
 & Platform and our data & Platform, but not our data & Neither platform nor our data \\ 
  \hline
1 & Unique applicant identifier & Full name & Athletic recruitment eligibility \\ 
  2 & Gender & High school transcript(s) & True admission outcome(s) \\ 
  3 & Race, ethnicity, and region(s) of origin & Academic honors & True enrollment outcome(s) \\ 
  4 & Age & Letters of recommendation & Ratings of admission officers \\ 
  5 & Citizenship status & Essays and written responses & Alumni interview ratings \\ 
  6 & High school name and location & Intended career & Official test scores \\ 
  7 & High school graduation date & College-specific fields (e.g., major) & Family income and assets \\ 
  8 & Self-reported test scores &  &  \\ 
  9 & Self-reported GPA, GPA weighting, and class rank &  &  \\ 
  10 & Highest educational attainment of parents &  &  \\ 
  11 & Institutions attended and degrees obtained by parents &  &  \\ 
  12 & Extracurricular categories, years participated, hours participated per year, leadership positions, and free text description &  &  \\ 
  13 & Application submission status at individual colleges &  &  \\ 
  14 & Application timing (e.g., restrictive early action) &  &  \\ 
  15 & Application fee waiver status at individual colleges &  &  \\ 
  16 & Receipt(s) of official transcript submission to individual colleges sent via the platform &  &  \\ 
   \hline
\caption{
    Variables observed by the national postsecondary application platform and the authors, only the platform, and neither the platform nor the authors.
} 
\label{fig:observed-unobserved-vars}
\end{longtable}

\restoregeometry

\FloatBarrier

\newpage

\begin{longtable}{|l|l|p{10cm}|}
  \hline
 & Covariate & Additional description \\ 
  \hline
1 & Intercept &  \\ 
  2 & South Asian & Applicant identifies as South Asian \\ 
  3 & Southeast Asian &  Applicant identifies as Southeast Asian\\ 
  4 & East Asian &  Applicant identifies as East Asian\\ 
  5 & Year by college fixed effects & Term for each combination of selective college applied to and application year, e.g., `College X 2016' \\ 
   \hline
\caption{
    Variables includes in Model 1, `Basket+year'. 
}  
\label{fig:basket-year-model-covariates}
\end{longtable}

\FloatBarrier

\newpage

\newgeometry{left=0.5in, right=0.5in}

\begin{longtable}{|l|l|p{10cm}|}
  \hline
 & Covariate & Additional description \\ 
  \hline
1 & Equivalent ACT Composite Score & If SAT score reported, converted to equivalent ACT score \\ 
  2 & Equivalent ACT Composite Score Squared &  \\ 
  3 & Missing ACT Score & Student did not report an ACT or SAT score \\ 
   \hline
\caption{
    Variables included in Model 2, `SAT/ACT'. Variables from Model 1 are also included. 
}  
\label{fig:sat-act-model-covariates}
\end{longtable}

\FloatBarrier

\newpage

\begin{longtable}{|l|l|p{10cm}|}
  \hline
 & Covariate & Additional description \\ 
  \hline
1 & Standardized GPA & GPA standardized by high school and year \\ 
  2 & Missing Cumulative GPA & Student did not report a GPA \\ 
  3 & Standardized ACT &  \\ 
  4 & Std. Num. AP & Standardized number of AP tests taken \\ 
  5 & Std. Num. Passed AP & Standardized number of AP tests with a reported score of 3 or higher \\ 
  6 & Std. Num. 5 AP & Standardized number of AP tests with a reported score of 5 (maximum) \\ 
  7 & Std. Num. SAT Subject &  \\ 
  8 & Std. Num. SAT Subject 700 & Standardized number of SAT subject tests with a score of at least 700 \\ 
  9 & Std. Num. Science AP &  \\ 
  10 & Std. Num. History AP &  \\ 
  11 & Std. Num. Math AP &  \\ 
  12 & Std. Num. English AP &  \\ 
  13 & Std. Num. Language AP &  \\ 
  14 & Std. Num. Social Science AP &  \\ 
  15 & Std. Num. Arts AP &  \\ 
  16 & Std. Num. Science SAT Subject &  \\ 
  17 & Std. Num. History SAT Subject &  \\ 
  18 & Std. Num. Math SAT Subject &  \\ 
  19 & Std. Num. English SAT Subject &  \\ 
  20 & Std. Num. Language SAT Subject &  \\ 
  21 & Took Art Studio Art 2D Design AP &  \\ 
  22 & Took Art Studio Art 3D Design AP &  \\ 
  23 & Took Art Studio Art Drawing AP &  \\ 
  24 & Took Biology AP &  \\ 
  25 & Took Biology Ecological SAT Subject &  \\ 
  26 & Took Biology Molecular SAT Subject &  \\ 
  27 & Took Calculus AB AP &  \\ 
  28 & Took Calculus BC AP &  \\ 
  29 & Took Calculus BC AB Subscore Grade AP & Reported a Calculus AB subscore for AP Calculus BC \\ 
  30 & Took Chemistry AP &  \\ 
  31 & Took Chemistry SAT Subject &  \\ 
  32 & Took Computer Science A AP &  \\ 
  33 & Took Economics Macroeconomics AP &  \\ 
  34 & Took Economics Microeconomics AP &  \\ 
  35 & Took English Language Composition AP &  \\ 
  36 & Took English Literature Composition AP &  \\ 
  37 & Took Environmental Science AP &  \\ 
  38 & Took European History AP &  \\ 
  39 & Took French Language AP &  \\ 
  40 & Took French Reading SAT Subject &  \\ 
  41 & Took French With Listening SAT Subject &  \\ 
  42 & Took German Language AP &  \\ 
  43 & Took German Reading SAT Subject &  \\ 
  44 & Took German With Listening SAT Subject &  \\ 
  45 & Took Government Politics Comparative AP &  \\ 
  46 & Took Government Politics United States AP &  \\ 
  47 & Took History Of Art AP &  \\ 
  48 & Took Human Geography AP &  \\ 
  49 & Took Italian Language Culture AP &  \\ 
  50 & Took Italian Reading SAT Subject &  \\ 
  51 & Took Latin AP &  \\ 
  52 & Took Latin Reading SAT Subject &  \\ 
  53 & Took Latin Literature AP &  \\ 
  54 & Took Latin Vergil AP &  \\ 
  55 & Took Literature SAT Subject &  \\ 
  56 & Took Math Level 1 SAT Subject &  \\ 
  57 & Took Math Level 2 SAT Subject &  \\ 
  58 & Took Music Theory AP &  \\ 
  59 & Took Music Theory Aural Subscore AP &  \\ 
  60 & Took Music Theory Nonaural Subscore AP &  \\ 
  61 & Took Physics SAT Subject &  \\ 
  62 & Took Physics 1 AP &  \\ 
  63 & Took Physics 2 AP &  \\ 
  64 & Took Physics B AP &  \\ 
  65 & Took Physics C Electricity Magnetism AP &  \\ 
  66 & Took Physics C Mechanics AP &  \\ 
  67 & Took Psychology AP &  \\ 
  68 & Took Research AP &  \\ 
  69 & Took Seminar AP &  \\ 
  70 & Took Spanish Language AP &  \\ 
  71 & Took Spanish Literature AP &  \\ 
  72 & Took Spanish Reading SAT Subject &  \\ 
  73 & Took Spanish With Listening SAT Subject &  \\ 
  74 & Took Statistics AP &  \\ 
  75 & Took US History SAT Subject &  \\ 
  76 & Took United States History AP &  \\ 
  77 & Took World History AP &  \\ 
  78 & Took World History SAT Subject &  \\ 
  79 & Took Writing SAT Subject &  \\ 
   \hline
\caption{
    Variables included in Model 3, `GPA+AP+SAT2'. Variables from all prior model are also included. Standardization is by high school-year using all applicants observed by the platform. We standardize by subtracting the sample mean and dividing by the sample standard deviation. Standardized values for high school-years with only one observation are coded as 0. Standardized values are capped at 3 and floored at -3.
}  
\label{fig:academics-model-covariates}
\end{longtable}

\FloatBarrier

\newpage

\begin{longtable}{|l|l|p{13cm}|}
  \hline
 & Covariate & Additional description \\ 
  \hline
1 & Archery & 8 covariates per sport: Log total number of hours participated in sport, binary indicator for leadership role in sport, binary indicator for four years of high school participation in sport, and binary indicator for leadership and four year participation in sport, with separate covariates for JV/Varsity participation and Club participation \\ 
  2 & Badminton &  \\ 
  3 & Baseball &  \\ 
  4 & Basketball &  \\ 
  5 & Bowling &  \\ 
  6 & Boxing &  \\ 
  7 & Cheerleading &  \\ 
  8 & Cricket &  \\ 
  9 & Crosscountry &  \\ 
  10 & Diving &  \\ 
  11 & Equestrian &  \\ 
  12 & Fencing &  \\ 
  13 & Field Hockey &  \\ 
  14 & Football &  \\ 
  15 & Golf &  \\ 
  16 & Gymnastics &  \\ 
  17 & Handball &  \\ 
  18 & Ice Hockey &  \\ 
  19 & Indoor Track &  \\ 
  20 & Judo &  \\ 
  21 & Lacrosse &  \\ 
  22 & Other Sport &  \\ 
  23 & Outdoor Track &  \\ 
  24 & Racquetball &  \\ 
  25 & Rifle &  \\ 
  26 & Rowing Crew &  \\ 
  27 & Rugby &  \\ 
  28 & Sailing &  \\ 
  29 & Skiing &  \\ 
  30 & Soccer &  \\ 
  31 & Softball &  \\ 
  32 & Squash &  \\ 
  33 & Swim &  \\ 
  34 & Sync swimming &  \\ 
  35 & Table Tennis &  \\ 
  36 & Tennis &  \\ 
  37 & Track and field &  \\ 
  38 & Triathlon &  \\ 
  39 & Volleyball &  \\ 
  40 & Water polo &  \\ 
  41 & Weight lifting &  \\ 
  42 & Wrestling &  \\ 
  43 & Academic & 4 covariates per activity type: Identical to sports, but without the JV/Varsity or Club designation \\ 
  44 & Art &  \\ 
  45 & Career Oriented &  \\ 
  46 & Volunteering &  \\ 
  47 & Computer/Technology &  \\ 
  48 & Cultural &  \\ 
  49 & Dance &  \\ 
  50 & Debate/Speech &  \\ 
  51 & Environmental &  \\ 
  52 & Family Responsibilities &  \\ 
  53 & Foreign Exchange &  \\ 
  54 & Foreign Language &  \\ 
  55 & Journalism/Publication &  \\ 
  56 & Junior ROTC &  \\ 
  57 & LGBT &  \\ 
  58 & Music Instrumental &  \\ 
  59 & Music Vocal &  \\ 
  60 & Other Activity &  \\ 
  61 & Religious &  \\ 
  62 & Research &  \\ 
  63 & Robotics &  \\ 
  64 & School Spirit &  \\ 
  65 & Science/Math &  \\ 
  66 & Student Govt/Politics &  \\ 
  67 & Theater/Drama &  \\ 
  68 & Work Paid &  \\ 
   \hline
\caption{
    Variables included in Model 4, `Activities'. Variables from all prior model are also included.
} 
\label{fig:activities-model-covariates}
\end{longtable}

\FloatBarrier

\newpage

\begin{longtable}{|l|l|p{9cm}|}
  \hline
 & Covariate & Additional description \\ 
  \hline
1 & Male &  \\ 
  2 & Received Platform Fee Waiver & Received an income-eligibility fee waiver for any school applied to via the platform \\ 
  3 & Received Subset Platform Fee Waiver & Received an income-eligbility fee waiver at any school among the selective schools we consider \\ 
  4 & Received Subset Member Fee Waiver & Received a fee waiver directly from the considered school (not necessarily related to income) \\ 
  5 & Highest Parent Educ. is High School &  \\ 
  6 & Highest Parent Educ. is Some College &  \\ 
  7 & Highest Parent Educ. is 4 year College Degree &  \\ 
  8 & Highest Parent Educ. is Graduate School &  \\ 
  9 & Highest Parent Educ. is Unknown &  \\ 
  10 & Top 50 Non Subset Legacy Undergrad 1 & First listed parent attended a Top 50 university defined by U.S. News in 2019 outside of the schools we consider as an undergraduate \\ 
  11 & Top 50 Non Subset Legacy Undergrad 2 &  \\ 
  12 & Top 50 Non Subset Legacy Grad 1 &  \\ 
  13 & Top 50 Non Subset Legacy Grad 2 &  \\ 
  14 & No App Subset Legacy Undergrad & Either parent was an undergraduate at a considered schools that the applicant did not apply to \\ 
  15 & No App Subset Legacy Grad &  \\ 
   \hline
\caption{
    Variables included in Model 5, `Sex+Family'. Variables from all prior model are also included.
} 
\label{fig:demographics-model-covariates}
\end{longtable}

\FloatBarrier

\newpage

\begin{longtable}{|l|l|p{10cm}|}
  \hline
 & Covariate & Additional description \\ 
  \hline
1 & Early Application Subset & Applied to a considered school under restrictive early action \\ 
  2 & Early Decision Subset &  \\ 
   \hline
\caption{
    Variables included in Model 6, `Early App'. Variables from Models 1 through 5 are also included. 
} 
\label{fig:early-app-model-covariates}
\end{longtable}

\FloatBarrier

\newpage

\begin{longtable}{|l|l|p{10cm}|}
  \hline
 & Covariate & Additional description \\ 
  \hline
1 & Subset Double Legacy Undergrad & Both parents were undergraduates at the same considered school to which the student applied \\ 
  2 & Subset Double Legacy Grad &  \\ 
  3 & Subset Double Legacy Mixed & One parent was an undergraduate and the other parent was a graduate student at the same considered school to which the student applied \\ 
  4 & Subset Single Legacy Undergrad & Exactly one parent was an undergraduate at a considered school to which the student applied \\ 
  5 & Subset Single Legacy Grad &  \\ 
  6 & Subset Two Separate Legacy Undergrad & Each parent was an undergraduate at a considered school to which the student applied and both attended a different considered school \\ 
  7 & Subset Two Separate Legacy Grad &  \\ 
   \hline
\caption{
    Variables included in Model 7, `Legacy'. Variables from Models 1 through 5 are also included. 
} 
\label{fig:legacy-model-covariates}
\end{longtable}

\FloatBarrier

\newpage

\begin{longtable}{|l|l|p{10cm}|}
  \hline
 & Covariate & Additional description \\ 
  \hline
1 & Log Graduating Class Size &  \\ 
  2 & Prop. Students Applying Platform & Proportion of students in the graduating class who submitted at least one application via the platform \\ 
  3 & Prop. Free Reduced Lunch &  \\ 
  4 & Missing Prop. Free Reduced Lunch & Unknown proportion of students receiving free or reduced lunch in high school \\ 
  5 & Is Private & Attended a private high school \\ 
  6 & Unknown Public/Private & Unknown classification of high school as public or private \\ 
  7 & Is Parochial & Attended a parochial high school \\ 
  8 & Top 100 Private & Top 100 private school according to 2022 Niche Rankings \\ 
  9 & Top 100 Public & Top 100 public school according to 2022 U.S. News rankings \\ 
  10 & School Offers AP/SAT2 Fixed Effects & For each of the AP and SAT subject tests identified above, did at least one applicant in the high school-year report a score for it? \\ 
  11 & Rurality & Terms for U.S. Census Rurality Code \\ 
  12 & ZIP3 Fixed Effects & Terms for first three digits of high school zip code \\ 
  13 & State-year-basket Fixed Effects & Terms for each combination of considered college applied to, high school state, and year of application, e.g., `College X 2016 California' \\ 
  14 & Log State ACT Rank & Logarithm of the within state-year ranking of applicant's ACT score \\ 
   \hline 
\caption{
    Variables included in Model 8, `Location+HS'. Variables from Models 1 through 5 are also included, except for basket-year fixed effects, which are redundant with the included state-year-basket fixed effects.
} 
\label{fig:location-model-covariates}
\end{longtable}

\FloatBarrier

\newpage

\restoregeometry

\setlength\LTleft{-2.5cm}

\begin{longtable}{|r|c|C{1cm}|C{1cm}|C{1cm}|C{1cm}|C{0.8cm}|C{1cm}|C{1cm}|C{1.2cm}|C{1cm}|C{0.8cm}|}
  \hline
Variant & Region & White base rate & Basket + year & SAT / ACT & GPA + AP + SAT2 & ECs & Sex + Fam. & Early app & Legacy & Loc. + HS & All \\ 
  \hline
Main model & E. Asian & 12\% & 1.11 & 0.86 & 0.85 & 0.83 & 0.79 & 0.73 & 0.90 & 0.88 & 0.90 \\ 
  E. Asian and white & E. Asian & 12\% & 1.12 & 0.87 & 0.84 & 0.81 & 0.78 & 0.73 & 0.89 & 0.88 & 0.89 \\ 
  Include recruits & E. Asian & 13.8\% & 1.08 & 0.85 & 0.83 & 0.86 & 0.83 & 0.76 & 0.92 & 0.91 & 0.90 \\ 
  2015 only & E. Asian & 12.8\% & 1.06 & 0.81 & 0.79 & 0.77 & 0.74 & 0.68 & 0.84 & 0.84 & 0.85 \\ 
  2016 only & E. Asian & 12.4\% & 1.10 & 0.82 & 0.82 & 0.78 & 0.74 & 0.68 & 0.85 & 0.81 & 0.83 \\ 
  2017 only & E. Asian & 11.9\% & 1.09 & 0.88 & 0.86 & 0.86 & 0.81 & 0.75 & 0.92 & 0.90 & 0.91 \\ 
  2018 only & E. Asian & 11.2\% & 1.15 & 0.91 & 0.88 & 0.85 & 0.82 & 0.76 & 0.92 & 0.92 & 0.94 \\ 
  2019 only & E. Asian & 11.9\% & 1.16 & 0.89 & 0.88 & 0.87 & 0.81 & 0.74 & 0.91 & 0.90 & 0.90 \\ 
  East coast only & E. Asian & 13.8\% & 1.20 & 0.93 & 0.88 & 0.86 & 0.82 & 0.73 & 0.94 & 0.87 & 0.87 \\ 
  California only & E. Asian & 10.5\% & 0.89 & 0.66 & 0.81 & 0.80 & 0.77 & 0.76 & 0.86 & 0.90 & 0.95 \\ 
  Real ACT/GPA & E. Asian & 11.5\% & 1.18 & 0.90 & 0.88 & 0.86 & 0.79 & 0.73 & 0.88 & 0.89 & 0.90 \\ 
  ACT $\geq$ 27 & E. Asian & 13.1\% & 1.12 & 0.86 & 0.84 & 0.82 & 0.76 & 0.71 & 0.86 & 0.87 & 0.89 \\ 
  Remove legacy & E. Asian & 9.7\% & 1.30 & 1.00 & 0.96 & 0.94 & 0.90 & 0.82 & 0.91 & 0.99 & 0.90 \\ 
  Transcript senders & E. Asian & 18.7\% & 1.36 & 1.06 & 0.99 & 0.95 & 0.89 & 0.82 & 1.00 & 0.92 & 0.93 \\ 
  Regular decision & E. Asian & 7.6\% & 0.97 & 0.77 & 0.80 & 0.79 & 0.73 & 0.71 & 0.81 & 0.81 & 0.87 \\ 
  No transcript thres. & E. Asian & 8.6\% & 1.19 & 0.90 & 0.87 & 0.85 & 0.82 & 0.76 & 0.92 & 0.90 & 0.91 \\ 
  20\% transcript thres. & E. Asian & 11.8\% & 1.12 & 0.87 & 0.85 & 0.84 & 0.80 & 0.74 & 0.90 & 0.89 & 0.90 \\ 
  0\% transcript thres. & E. Asian & 12.2\% & 1.11 & 0.86 & 0.84 & 0.83 & 0.79 & 0.72 & 0.89 & 0.88 & 0.89 \\ 
  Reweighted & E. Asian & 12\% & 1.15 & 0.88 & 0.86 & 0.85 & 0.79 & 0.73 & 0.88 & 0.90 & 0.91 \\ 
  \hline
  Leave one out max & E. Asian & 12.1\% & 1.13 & 0.88 & 0.87 & 0.86 & 0.81 & 0.75 & 0.91 & 0.89 & 0.91 \\
  Leave one out min & E. Asian & 10.9\% & 1.05 & 0.81 & 0.81 & 0.79 & 0.75 & 0.71 & 0.85 & 0.83 & 0.87 \\
\hline
\caption{
    Robustness checks of the main specification.
    Each variant of the main specification lists the corresponding value of the exponentiated East Asian coefficient for each of the nine models in the main analysis. 
    Exponeniated coefficients are qualitatively similar across all specifications. 
    \textbf{Detailed descriptions of each variant are on the next page.}
} 
\label{fig:robustness-east-asian}
\end{longtable}

\FloatBarrier

\newpage

Detailed descriptions of each model variant:
\begin{itemize}
        \item The `E. Asian and white' variant fits the main specification only on East Asian and white applicants in the study pool, mimicking the effect of interacting race with each variable in the main model.
        \item The `Include recruits' specification does not remove applicants who we believe may be recruited athletes.
        \item The `2015 only' model fits the main model on only the 2015-2016 academic year application data, with a similar interpretation for the other variants whose names end in `only'. 
        \item The `Real ACT/GPA' model excludes applicants who do not report an ACT/SAT score and/or a high school GPA.
        \item The `ACT $\geq$ 27' model removes applicants with an equivalent ACT below 27, as very few enrollees at the schools we consider have ACT scores below 27.
        \item The `Remove legacy' model removes legacy applicants from the study pool, following a similar model choice in the Harvard v.\ SFFA court case.
        \item The `Transcript senders' model includes only those applicants who sent a transcript to a specific college on the platform. 
        These applicants have the strongest enrollment signal, as the precision of our transcript heuristic is 97\%.
        \item The `Regular decision' model excludes applicants who applied early to only one college, sent a transcript to that college, and did not apply anywhere else. This is a likely signal of enrollment at the school to which the student applied early. 
        \item The `No transcript thres.' allows all high school-years to be included in the analysis, and only excludes students with at least one application who sent the same number of transcripts as applications.
        The `20\% transcript thres.' model allows only applicants from high school-years for which less than 20\% of applicants who submitted more than one application sent the same number of transcripts and applications.
        This model also removes all students with more than one application who sent the same number of transcripts as applications.
        The `0\% transcript thres.' model does not allow high-school years with any applicants who submitted more than one application and sent the same number of transcripts and applications.
        \item The `Reweighted' model reweights the main model by the inverse likelihood that the given applicant attended a high school-year where no more than 5\% of applicants with more than one application sent the same number of transcripts as applications.
        The corresponding propensity model is fit using the same covariates as Model 9, excluding the state-year-basket fixed effects. 
        \item The `Leave one out' variants assess the sensitivity of the Asian region coefficients to the particular set of schools considered in the analysis.
        The exponentiated coefficients of the `Leave one out max' and `Leave one out min' variants are derived from fitting the 9 models $n_{\text{schools}}$ times, where $n_{\text{schools}}$ is the number of selective schools considered in the main analysis. 
        To preserve confidentiality, the exact value of $n_{\text{schools}}$ is not provided.
        For each of the 9 model specifications, we report the maximum and minimum observed values of the exponentiated Asian region coefficient across $n_{\text{schools}}$ datasets. 
        Each dataset excludes application data from one of the selective schools considered in the main analysis. 
    \end{itemize}

\FloatBarrier

\newpage

\begin{longtable}{|r|c|C{1cm}|C{1cm}|C{1cm}|C{1cm}|C{0.8cm}|C{1cm}|C{1cm}|C{1.2cm}|C{1cm}|C{0.8cm}|}
  \hline
Variant & Region & White base rate & Basket + year & SAT / ACT & GPA + AP + SAT2 & ECs & Sex + Fam. & Early app & Legacy & Loc. + HS & All \\ 
  \hline
Main model & S. Asian & 12\% & 0.66 & 0.56 & 0.59 & 0.51 & 0.51 & 0.52 & 0.61 & 0.60 & 0.70 \\ 
  S. Asian and white & S. Asian & 12\% & 0.64 & 0.55 & 0.58 & 0.49 & 0.49 & 0.51 & 0.60 & 0.60 & 0.71 \\ 
  Include recruits & S. Asian & 13.8\% & 0.62 & 0.53 & 0.55 & 0.52 & 0.52 & 0.53 & 0.61 & 0.61 & 0.69 \\ 
  2015 only & S. Asian & 12.8\% & 0.63 & 0.53 & 0.57 & 0.50 & 0.49 & 0.50 & 0.59 & 0.59 & 0.69 \\ 
  2016 only & S. Asian & 12.4\% & 0.65 & 0.53 & 0.57 & 0.49 & 0.48 & 0.50 & 0.58 & 0.56 & 0.67 \\ 
  2017 only & S. Asian & 11.9\% & 0.67 & 0.58 & 0.62 & 0.53 & 0.53 & 0.54 & 0.63 & 0.62 & 0.71 \\ 
  2018 only & S. Asian & 11.2\% & 0.65 & 0.56 & 0.59 & 0.51 & 0.51 & 0.52 & 0.61 & 0.59 & 0.68 \\ 
  2019 only & S. Asian & 11.9\% & 0.69 & 0.58 & 0.62 & 0.53 & 0.52 & 0.52 & 0.61 & 0.61 & 0.68 \\ 
  East coast only & S. Asian & 13.8\% & 0.64 & 0.56 & 0.59 & 0.51 & 0.50 & 0.51 & 0.60 & 0.59 & 0.68 \\ 
  California only & S. Asian & 10.5\% & 0.66 & 0.49 & 0.64 & 0.57 & 0.55 & 0.60 & 0.68 & 0.65 & 0.79 \\ 
  Real ACT/GPA & S. Asian & 11.5\% & 0.70 & 0.58 & 0.61 & 0.53 & 0.51 & 0.53 & 0.60 & 0.61 & 0.69 \\ 
  ACT $\geq$ 27 & S. Asian & 13.1\% & 0.68 & 0.56 & 0.60 & 0.52 & 0.50 & 0.52 & 0.60 & 0.60 & 0.69 \\ 
  Remove legacy & S. Asian & 9.7\% & 0.78 & 0.65 & 0.68 & 0.59 & 0.60 & 0.60 & 0.61 & 0.70 & 0.69 \\ 
  Transcript senders & S. Asian & 18.7\% & 0.81 & 0.69 & 0.69 & 0.58 & 0.57 & 0.58 & 0.68 & 0.64 & 0.73 \\ 
  Regular decision & S. Asian & 7.6\% & 0.60 & 0.52 & 0.58 & 0.53 & 0.51 & 0.52 & 0.60 & 0.59 & 0.68 \\ 
  No transcript thres. & S. Asian & 8.6\% & 0.69 & 0.58 & 0.60 & 0.53 & 0.53 & 0.55 & 0.63 & 0.62 & 0.71 \\ 
  20\% transcript thres. & S. Asian & 11.8\% & 0.66 & 0.56 & 0.60 & 0.52 & 0.51 & 0.53 & 0.61 & 0.61 & 0.70 \\ 
  0\% transcript thres. & S. Asian & 12.2\% & 0.67 & 0.57 & 0.60 & 0.52 & 0.52 & 0.53 & 0.62 & 0.62 & 0.70 \\ 
  Reweighted & S. Asian & 12\% & 0.67 & 0.56 & 0.59 & 0.52 & 0.51 & 0.53 & 0.60 & 0.62 & 0.71 \\ 
  \hline
  Leave one out max & S. Asian & 12.1\% & 0.67 & 0.57 & 0.60 & 0.52 & 0.51 & 0.53 & 0.62 & 0.61 & 0.71 \\
  Leave one out min & S Asian & 10.9\% & 0.64 & 0.53 & 0.57 & 0.51 & 0.50 & 0.52 & 0.59 & 0.59 & 0.68 \\
\hline
\caption{
    Robustness checks of the main specification.
    Each variant of the main specification lists the corresponding value of the exponentiated South Asian coefficient for each of the nine models in the main analysis. 
    Exponeniated coefficients are qualitatively similar across all specifications. 
     Model variants are described in the caption of the corresponding figure for the exponentiated East Asian coefficient.
} 
\label{fig:robustness-south-asian}
\end{longtable}

\FloatBarrier

\newpage

\begin{longtable}
{|r|c|C{1cm}|C{1cm}|C{1cm}|C{1cm}|C{0.8cm}|C{1cm}|C{1cm}|C{1.2cm}|C{1cm}|C{0.8cm}|}
  \hline
Variant & Region & White base rate & Basket + year & SAT / ACT & GPA + AP + SAT2 & ECs & Sex + Fam. & Early app & Legacy & Loc. + HS & All \\ 
  \hline
Main model & SE Asian & 12\% & 0.64 & 0.73 & 0.78 & 0.83 & 0.81 & 0.84 & 0.88 & 0.94 & 1.02 \\ 
  SE Asian and white & SE Asian & 12\% & 0.64 & 0.73 & 0.76 & 0.80 & 0.78 & 0.82 & 0.85 & 0.90 & 0.98 \\ 
  Include recruits & SE Asian & 13.8\% & 0.60 & 0.69 & 0.72 & 0.81 & 0.80 & 0.83 & 0.87 & 0.92 & 0.99 \\ 
  2015 only & SE Asian & 12.8\% & 0.64 & 0.76 & 0.79 & 0.83 & 0.82 & 0.88 & 0.90 & 1.01 & 1.15 \\ 
  2016 only & SE Asian & 12.4\% & 0.62 & 0.71 & 0.75 & 0.82 & 0.78 & 0.79 & 0.86 & 0.91 & 0.98 \\ 
  2017 only & SE Asian & 11.9\% & 0.64 & 0.74 & 0.79 & 0.84 & 0.83 & 0.85 & 0.91 & 0.97 & 1.04 \\ 
  2018 only & SE Asian & 11.2\% & 0.64 & 0.72 & 0.77 & 0.82 & 0.80 & 0.85 & 0.87 & 0.94 & 1.06 \\ 
  2019 only & SE Asian & 11.9\% & 0.67 & 0.75 & 0.80 & 0.84 & 0.79 & 0.81 & 0.85 & 0.92 & 0.99 \\ 
  East coast only & SE Asian & 13.8\% & 0.63 & 0.73 & 0.76 & 0.81 & 0.79 & 0.80 & 0.86 & 0.84 & 0.91 \\ 
  California only & SE Asian & 10.5\% & 0.60 & 0.65 & 0.84 & 0.90 & 0.88 & 0.95 & 0.98 & 1.04 & 1.17 \\ 
  Real ACT/GPA & SE Asian & 11.5\% & 0.69 & 0.81 & 0.85 & 0.88 & 0.82 & 0.85 & 0.88 & 0.96 & 1.03 \\ 
  ACT $\geq$ 27 & SE Asian & 13.1\% & 0.68 & 0.75 & 0.81 & 0.85 & 0.80 & 0.84 & 0.87 & 0.95 & 1.04 \\ 
  Remove legacy & SE Asian & 9.7\% & 0.76 & 0.86 & 0.90 & 0.95 & 0.90 & 0.92 & 0.91 & 1.04 & 1.04 \\ 
  Transcript senders & SE Asian & 18.7\% & 0.85 & 0.98 & 1.02 & 1.05 & 1.00 & 1.04 & 1.10 & 1.02 & 1.11 \\ 
  Regular decision & SE Asian & 7.6\% & 0.70 & 0.79 & 0.86 & 0.91 & 0.84 & 0.85 & 0.91 & 0.97 & 1.05 \\ 
  No transcript thres. & SE Asian & 8.6\% & 0.62 & 0.72 & 0.76 & 0.82 & 0.81 & 0.85 & 0.88 & 0.95 & 1.04 \\ 
  20\% transcript thres. & SE Asian & 11.8\% & 0.65 & 0.74 & 0.79 & 0.84 & 0.82 & 0.85 & 0.89 & 0.95 & 1.03 \\ 
  0\% transcript thres. & SE Asian & 12.2\% & 0.64 & 0.74 & 0.79 & 0.84 & 0.81 & 0.85 & 0.88 & 0.94 & 1.02 \\ 
  Reweighted & SE Asian & 12\% & 0.71 & 0.83 & 0.88 & 0.94 & 0.88 & 0.91 & 0.94 & 1.05 & 1.13 \\ 
  \hline
  Leave one out max & SE Asian & 12.1\% & 0.66 & 0.76 & 0.82 & 0.88 & 0.84 & 0.88 & 0.92 & 0.95 & 1.05 \\
  Leave one out min & SE Asian & 10.9\% & 0.58 & 0.64 & 0.68 & 0.72 & 0.69 & 0.75 & 0.75 & 0.82 & 0.90 \\
\hline
\caption{
    Robustness checks of the main specification.
    Each variant of the main specification lists the corresponding value of the exponentiated Southeast Asian coefficient for each of the nine models in the main analysis. 
    Exponeniated coefficients are qualitatively similar across all specifications. 
    Model variants are described in the caption of the corresponding figure for the exponentiated East Asian coefficient.
} 
\label{fig:robustness-southeast-asian}
\end{longtable}

\FloatBarrier

\newpage

\restoregeometry

\setlength\LTleft{-1.5cm}

\newgeometry{top=0.3in, bottom=0.5in}

\begin{longtable}{|c|c|c||c|c||c|c|}
 \hline Race & ZIP 1 & ACT & Non-legacy applicants & Non-legacy admits & Legacy applicants & Legacy admits \\ \hline
White & 0 & 32 & 3,321 & 199 & 411 & 102 \\ 
   \hline
White & 0 & 33 & 4,652 & 409 & 705 & 166 \\ 
   \hline
White & 0 & 34 & 5,201 & 746 & 942 & 357 \\ 
   \hline
White & 0 & 35 & 4,380 & 1,058 & 980 & 449 \\ 
   \hline
White & 0 & 36 & 1,600 & 609 & 451 & 306 \\ 
   \hline
White & 1 & 32 & 3,236 & 286 & 285 & 77 \\ 
   \hline
White & 1 & 33 & 4,455 & 516 & 622 & 214 \\ 
   \hline
White & 1 & 34 & 5,023 & 1,002 & 849 & 373 \\ 
   \hline
White & 1 & 35 & 4,093 & 1,198 & 898 & 538 \\ 
   \hline
White & 1 & 36 & 1,374 & 555 & 404 & 295 \\ 
   \hline
White & 2 & 32 & 1,618 & 77 & 155 & 21 \\ 
   \hline
White & 2 & 33 & 2,408 & 159 & 252 & 33 \\ 
   \hline
White & 2 & 34 & 2,770 & 285 & 385 & 119 \\ 
   \hline
White & 2 & 35 & 2,298 & 461 & 410 & 177 \\ 
   \hline
White & 2 & 36 & 868 & 252 & 184 & 108 \\ 
   \hline
White & 3 & 32 & 1,148 & 49 & 59 & 14 \\ 
   \hline
White & 3 & 33 & 1,630 & 111 & 103 & 23 \\ 
   \hline
White & 3 & 34 & 1,949 & 207 & 157 & 47 \\ 
   \hline
White & 3 & 35 & 1,717 & 253 & 151 & 74 \\ 
   \hline
White & 3 & 36 & 708 & 191 & 78 & 45 \\ 
   \hline
White & 4 & 32 & 1,022 & 37 &  &  \\ 
   \hline
White & 4 & 33 & 1,400 & 78 & 77 & 13 \\ 
   \hline
White & 4 & 34 & 1,521 & 128 & 82 & 29 \\ 
   \hline
White & 4 & 35 & 1,364 & 185 & 81 & 32 \\ 
   \hline
White & 4 & 36 & 464 & 78 &  &  \\ 
   \hline
White & 5 & 32 & 548 & 22 &  &  \\ 
   \hline
White & 5 & 33 & 768 & 59 &  &  \\ 
   \hline
White & 5 & 34 & 821 & 55 & 54 & 16 \\ 
   \hline
White & 5 & 35 & 745 & 111 & 65 & 31 \\ 
   \hline
White & 5 & 36 & 270 & 64 &  &  \\ 
   \hline
White & 6 & 32 & 1,082 & 52 & 65 & 17 \\ 
   \hline
White & 6 & 33 & 1,606 & 100 & 95 & 19 \\ 
   \hline
White & 6 & 34 & 1,890 & 193 & 156 & 52 \\ 
   \hline
White & 6 & 35 & 1,729 & 258 & 135 & 43 \\ 
   \hline
White & 6 & 36 & 716 & 132 & 75 & 42 \\ 
   \hline
White & 7 & 32 & 692 & 31 &  &  \\ 
   \hline
White & 7 & 33 & 924 & 36 &  &  \\ 
   \hline
White & 7 & 34 & 1,263 & 106 & 78 & 21 \\ 
   \hline
White & 7 & 35 & 1,143 & 148 & 85 & 35 \\ 
   \hline
White & 7 & 36 & 494 & 105 & 54 & 31 \\ 
   \hline
White & 8 & 32 & 786 & 36 & 59 & 8 \\ 
   \hline
White & 8 & 33 & 1,052 & 72 & 83 & 19 \\ 
   \hline
White & 8 & 34 & 1,148 & 117 & 101 & 36 \\ 
   \hline
White & 8 & 35 & 995 & 152 & 91 & 34 \\ 
   \hline
White & 8 & 36 & 334 & 76 & 53 & 29 \\ 
   \hline
White & 9 & 32 & 2,418 & 107 & 285 & 47 \\ 
   \hline
White & 9 & 33 & 3,396 & 195 & 431 & 88 \\ 
   \hline
White & 9 & 34 & 3,969 & 416 & 583 & 175 \\ 
   \hline
White & 9 & 35 & 3,298 & 530 & 584 & 217 \\ 
   \hline
White & 9 & 36 & 1,108 & 276 & 243 & 127 \\ 
   \hline
\caption{
    Aggregated counts of white applicants and admits across groups defined by geography, equivalent ACT score, and legacy status.
    Admission is proxied by observing whether a final transcript is sent to one of the schools we consider.
    ``ZIP 1'' refers to the first digit of the student's high school ZIP code.
    To preserve confidentiality, legacy and non-legacy applicant cell counts with fewer than 50 applicants are redacted, along with the corresponding count of admits.
    Further, legacy and non-legacy admit cell counts of 0 are redacted, along with the corresponding count of applicants.
} 
\label{fig:replication_data_white}
\end{longtable}

\FloatBarrier

\newpage

\setlength\LTleft{-2cm}

\begin{longtable}{|c|c|c||c|c||c|c|}
 \hline
 Ethnicity & ZIP 1 & ACT & Non-legacy applicants & Non-legacy admits & Legacy applicants & Legacy admits \\ 
  \hline
 Asian & 0 & 32 & 1,109 & 56 & 51 & 12 \\ 
   \hline
Asian & 0 & 33 & 1,988 & 146 & 86 & 20 \\ 
   \hline
Asian & 0 & 34 & 3,050 & 353 & 181 & 50 \\ 
   \hline
Asian & 0 & 35 & 4,037 & 948 & 209 & 104 \\ 
   \hline
Asian & 0 & 36 & 2,538 & 1,027 & 86 & 56 \\ 
   \hline
Asian & 1 & 32 & 1,596 & 104 &  &  \\ 
   \hline
Asian & 1 & 33 & 2,363 & 265 & 73 & 27 \\ 
   \hline
Asian & 1 & 34 & 3,126 & 487 & 115 & 55 \\ 
   \hline
Asian & 1 & 35 & 3,204 & 985 & 165 & 85 \\ 
   \hline
Asian & 1 & 36 & 1,569 & 733 & 90 & 63 \\ 
   \hline
Asian & 2 & 32 & 628 & 19 &  &  \\ 
   \hline
Asian & 2 & 33 & 1,067 & 44 &  &  \\ 
   \hline
Asian & 2 & 34 & 1,513 & 154 & 51 & 8 \\ 
   \hline
Asian & 2 & 35 & 1,996 & 432 & 97 & 51 \\ 
   \hline
Asian & 2 & 36 & 1,103 & 358 &  &  \\ 
   \hline
Asian & 3 & 32 & 432 & 9 &  &  \\ 
   \hline
Asian & 3 & 33 & 754 & 41 &  &  \\ 
   \hline
Asian & 3 & 34 & 1,147 & 95 &  &  \\ 
   \hline
Asian & 3 & 35 & 1,347 & 227 &  &  \\ 
   \hline
Asian & 3 & 36 & 887 & 297 &  &  \\ 
   \hline
Asian & 4 & 32 & 269 & 10 &  &  \\ 
   \hline
Asian & 4 & 33 & 434 & 23 &  &  \\ 
   \hline
Asian & 4 & 34 & 730 & 72 &  &  \\ 
   \hline
Asian & 4 & 35 & 915 & 172 &  &  \\ 
   \hline
Asian & 4 & 36 & 604 & 162 &  &  \\ 
   \hline
Asian & 5 & 32 & 143 & 7 &  &  \\ 
   \hline
Asian & 5 & 33 & 193 & 15 &  &  \\ 
   \hline
Asian & 5 & 34 & 288 & 30 &  &  \\ 
   \hline
Asian & 5 & 35 & 375 & 73 &  &  \\ 
   \hline
Asian & 5 & 36 & 233 & 85 &  &  \\ 
   \hline
Asian & 6 & 32 & 412 & 10 &  &  \\ 
   \hline
Asian & 6 & 33 & 616 & 33 &  &  \\ 
   \hline
Asian & 6 & 34 & 876 & 77 &  &  \\ 
   \hline
Asian & 6 & 35 & 1,125 & 162 &  &  \\ 
   \hline
Asian & 6 & 36 & 739 & 231 &  &  \\ 
   \hline
Asian & 7 & 32 & 321 & 9 &  &  \\ 
   \hline
Asian & 7 & 33 & 604 & 27 &  &  \\ 
   \hline
Asian & 7 & 34 & 987 & 80 &  &  \\ 
   \hline
Asian & 7 & 35 & 1,529 & 213 &  &  \\ 
   \hline
Asian & 7 & 36 & 1,040 & 250 &  &  \\ 
   \hline
Asian & 8 & 32 & 222 & 10 &  &  \\ 
   \hline
Asian & 8 & 33 & 327 & 21 &  &  \\ 
   \hline
Asian & 8 & 34 & 512 & 58 &  &  \\ 
   \hline
Asian & 8 & 35 & 553 & 128 &  &  \\ 
   \hline
Asian & 8 & 36 & 344 & 122 &  &  \\ 
   \hline
Asian & 9 & 32 & 2,100 & 70 & 68 & 4 \\ 
   \hline
Asian & 9 & 33 & 3,473 & 163 & 126 & 25 \\ 
   \hline
Asian & 9 & 34 & 5,348 & 379 & 205 & 56 \\ 
   \hline
Asian & 9 & 35 & 6,692 & 877 & 255 & 83 \\ 
   \hline
Asian & 9 & 36 & 3,523 & 771 & 132 & 59 \\ 
   \hline
\caption{
    Aggregated counts of Asian American applicants and admits across groups defined by geography, equivalent ACT score, and legacy status.
    ``ZIP 1'' refers to the first digit of the student's high school ZIP code.
    To preserve confidentiality, legacy and non-legacy applicant cell counts with fewer than 50 applicants are redacted, along with the corresponding count of admits.
    Further, legacy and non-legacy admit cell counts of 0 are redacted, along with the corresponding count of applicants.
} 
\label{fig:replication_data_asian}
\end{longtable}

\FloatBarrier

\newpage

\vspace{-10em}
\begin{longtable}{|c|c|c||c|c||c|c|}
 \hline Ethnicity & ZIP 1 & ACT & Non-legacy applicants & Non-legacy admits & Legacy applicants & Legacy admits \\ \hline
  South Asian & 0 & 32 & 422 & 8 &  &  \\ 
   \hline
South Asian & 0 & 33 & 883 & 42 &  &  \\ 
   \hline
South Asian & 0 & 34 & 1,314 & 106 &  &  \\ 
   \hline
South Asian & 0 & 35 & 1,561 & 277 &  &  \\ 
   \hline
South Asian & 0 & 36 & 894 & 285 &  &  \\ 
   \hline
South Asian & 1 & 32 & 516 & 28 &  &  \\ 
   \hline
South Asian & 1 & 33 & 737 & 65 &  &  \\ 
   \hline
South Asian & 1 & 34 & 968 & 148 &  &  \\ 
   \hline
South Asian & 1 & 35 & 928 & 244 &  &  \\ 
   \hline
South Asian & 1 & 36 & 366 & 155 &  &  \\ 
   \hline
South Asian & 2 & 32 & 264 & 4 &  &  \\ 
   \hline
South Asian & 2 & 33 & 460 & 21 &  &  \\ 
   \hline
South Asian & 2 & 34 & 615 & 47 &  &  \\ 
   \hline
South Asian & 2 & 35 & 808 & 135 &  &  \\ 
   \hline
South Asian & 2 & 36 & 372 & 100 &  &  \\ 
   \hline
South Asian & 3 & 32 & 185 & 4 &  &  \\ 
   \hline
South Asian & 3 & 33 & 363 & 14 &  &  \\ 
   \hline
South Asian & 3 & 34 & 551 & 32 &  &  \\ 
   \hline
South Asian & 3 & 35 & 594 & 86 &  &  \\ 
   \hline
South Asian & 3 & 36 & 365 & 101 &  &  \\ 
   \hline
South Asian & 4 & 32 & 128 & 3 &  &  \\ 
   \hline
South Asian & 4 & 33 & 198 & 8 &  &  \\ 
   \hline
South Asian & 4 & 34 & 352 & 33 &  &  \\ 
   \hline
South Asian & 4 & 35 & 424 & 70 &  &  \\ 
   \hline
South Asian & 4 & 36 & 240 & 48 &  &  \\ 
   \hline
South Asian & 5 & 32 & 55 & 3 &  &  \\ 
   \hline
South Asian & 5 & 33 & 79 & 3 &  &  \\ 
   \hline
South Asian & 5 & 34 & 117 & 12 &  &  \\ 
   \hline
South Asian & 5 & 35 & 152 & 28 &  &  \\ 
   \hline
South Asian & 5 & 36 & 60 & 17 &  &  \\ 
   \hline
South Asian & 6 & 32 & 166 & 4 &  &  \\ 
   \hline
South Asian & 6 & 33 & 260 & 9 &  &  \\ 
   \hline
South Asian & 6 & 34 & 387 & 25 &  &  \\ 
   \hline
South Asian & 6 & 35 & 483 & 54 &  &  \\ 
   \hline
South Asian & 6 & 36 & 272 & 66 &  &  \\ 
   \hline
South Asian & 7 & 32 & 142 & 4 &  &  \\ 
   \hline
South Asian & 7 & 33 & 287 & 11 &  &  \\ 
   \hline
South Asian & 7 & 34 & 503 & 32 &  &  \\ 
   \hline
South Asian & 7 & 35 & 707 & 84 &  &  \\ 
   \hline
South Asian & 7 & 36 & 444 & 74 &  &  \\ 
   \hline
South Asian & 8 & 32 & 60 & 3 &  &  \\ 
   \hline
South Asian & 8 & 33 & 110 & 4 &  &  \\ 
   \hline
South Asian & 8 & 34 & 172 & 16 &  &  \\ 
   \hline
South Asian & 8 & 35 & 196 & 40 &  &  \\ 
   \hline
South Asian & 8 & 36 & 132 & 45 &  &  \\ 
   \hline
South Asian & 9 & 32 & 434 & 16 &  &  \\ 
   \hline
South Asian & 9 & 33 & 781 & 24 &  &  \\ 
   \hline
South Asian & 9 & 34 & 1,232 & 61 &  &  \\ 
   \hline
South Asian & 9 & 35 & 1,791 & 159 &  &  \\ 
   \hline
South Asian & 9 & 36 & 997 & 195 &  &  \\ 
   \hline
\caption{
    Aggregated counts of South Asian applicants and admits across groups defined by geography, equivalent ACT score, and legacy status.
    Admission is proxied by observing whether a final transcript is sent to one of the schools we consider.
    ``ZIP 1'' refers to the first digit of the student's high school ZIP code.
    To preserve confidentiality, legacy and non-legacy applicant cell counts with fewer than 50 applicants are redacted, along with the corresponding count of admits.
    Further, legacy and non-legacy admit cell counts of 0 are redacted, along with the corresponding count of applicants.
}  
\label{fig:replication_data_south_asian}
\end{longtable}

\FloatBarrier

\newpage

\setlength\LTleft{-2.5cm}

\begin{longtable}{|c|c|c||c|c||c|c|}
 \hline Ethnicity & ZIP 1 & ACT & Non-legacy applicants & Non-legacy admits & Legacy applicants & Legacy admits \\ \hline
Southeast Asian & 0 & 32 & 189 & 7 &  &  \\ 
   \hline
Southeast Asian & 0 & 33 & 253 & 21 &  &  \\ 
   \hline
Southeast Asian & 0 & 34 & 270 & 37 &  &  \\ 
   \hline
Southeast Asian & 0 & 35 & 263 & 69 &  &  \\ 
   \hline
Southeast Asian & 0 & 36 & 82 & 30 &  &  \\ 
   \hline
Southeast Asian & 1 & 32 & 208 & 9 &  &  \\ 
   \hline
Southeast Asian & 1 & 33 & 298 & 45 &  &  \\ 
   \hline
Southeast Asian & 1 & 34 & 293 & 41 &  &  \\ 
   \hline
Southeast Asian & 1 & 35 & 217 & 53 &  &  \\ 
   \hline
Southeast Asian & 1 & 36 & 92 & 38 &  &  \\ 
   \hline
Southeast Asian & 2 & 32 & 105 & 5 &  &  \\ 
   \hline
Southeast Asian & 2 & 33 & 145 & 3 &  &  \\ 
   \hline
Southeast Asian & 2 & 34 & 157 & 15 &  &  \\ 
   \hline
Southeast Asian & 2 & 35 & 135 & 24 &  &  \\ 
   \hline
Southeast Asian & 2 & 36 & 52 & 16 &  &  \\ 
   \hline
Southeast Asian & 3 & 32 & 78 & 2 &  &  \\ 
   \hline
Southeast Asian & 3 & 33 & 112 & 11 &  &  \\ 
   \hline
Southeast Asian & 3 & 34 & 128 & 16 &  &  \\ 
   \hline
Southeast Asian & 3 & 35 & 115 & 23 &  &  \\ 
   \hline
Southeast Asian & 3 & 36 &  &  &  &  \\ 
   \hline
Southeast Asian & 4 & 32 &  &  &  &  \\ 
   \hline
Southeast Asian & 4 & 33 & 57 & 4 &  &  \\ 
   \hline
Southeast Asian & 4 & 34 & 52 & 4 &  &  \\ 
   \hline
Southeast Asian & 4 & 35 & 57 & 12 &  &  \\ 
   \hline
Southeast Asian & 4 & 36 &  &  &  &  \\ 
   \hline
Southeast Asian & 5 & 32 &  &  &  &  \\ 
   \hline
Southeast Asian & 5 & 33 &  &  &  &  \\ 
   \hline
Southeast Asian & 5 & 34 &  &  &  &  \\ 
   \hline
Southeast Asian & 5 & 35 &  &  &  &  \\ 
   \hline
Southeast Asian & 5 & 36 &  &  &  &  \\ 
   \hline
Southeast Asian & 6 & 32 & 71 & 3 &  &  \\ 
   \hline
Southeast Asian & 6 & 33 & 108 & 7 &  &  \\ 
   \hline
Southeast Asian & 6 & 34 & 98 & 9 &  &  \\ 
   \hline
Southeast Asian & 6 & 35 & 80 & 9 &  &  \\ 
   \hline
Southeast Asian & 6 & 36 &  &  &  &  \\ 
   \hline
Southeast Asian & 7 & 32 & 70 & 2 &  &  \\ 
   \hline
Southeast Asian & 7 & 33 & 117 & 5 &  &  \\ 
   \hline
Southeast Asian & 7 & 34 & 109 & 9 &  &  \\ 
   \hline
Southeast Asian & 7 & 35 & 171 & 21 &  &  \\ 
   \hline
Southeast Asian & 7 & 36 & 88 & 18 &  &  \\ 
   \hline
Southeast Asian & 8 & 32 & 52 & 3 &  &  \\ 
   \hline
Southeast Asian & 8 & 33 & 53 & 5 &  &  \\ 
   \hline
Southeast Asian & 8 & 34 & 68 & 11 &  &  \\ 
   \hline
Southeast Asian & 8 & 35 & 58 & 11 &  &  \\ 
   \hline
Southeast Asian & 8 & 36 &  &  &  &  \\ 
   \hline
Southeast Asian & 9 & 32 & 470 & 17 &  &  \\ 
   \hline
Southeast Asian & 9 & 33 & 648 & 38 &  &  \\ 
   \hline
Southeast Asian & 9 & 34 & 814 & 56 &  &  \\ 
   \hline
Southeast Asian & 9 & 35 & 771 & 119 &  &  \\ 
   \hline
Southeast Asian & 9 & 36 & 249 & 55 &  &  \\ 
\hline
\caption{
    Aggregated counts of Southeast Asian applicants and admits across groups defined by geography, equivalent ACT score, and legacy status.
    Admission is proxied by observing whether a final transcript is sent to one of the schools we consider.
    "ZIP 1" refers to the first digit of the student's high school ZIP code.
    To preserve confidentiality, legacy and non-legacy applicant cell counts with fewer than 50 applicants are redacted, along with the corresponding count of admits.
    Further, legacy and non-legacy admit cell counts of 0 are redacted, along with the corresponding count of applicants.
}  
\label{fig:replication_data_southeast_asian}
\end{longtable}

\FloatBarrier

\newpage

\setlength\LTleft{-2cm}

\begin{longtable}{|c|c|c||c|c||c|c|}
 \hline Ethnicity & ZIP 1 & ACT & Non-legacy applicants & Non-legacy admits & Legacy applicants & Legacy admits \\ \hline
 East Asian & 0 & 32 & 498 & 41 &  &  \\ 
   \hline
East Asian & 0 & 33 & 852 & 83 & 61 & 18 \\ 
   \hline
East Asian & 0 & 34 & 1,466 & 210 & 124 & 34 \\ 
   \hline
East Asian & 0 & 35 & 2,213 & 602 & 156 & 81 \\ 
   \hline
East Asian & 0 & 36 & 1,562 & 712 & 65 & 47 \\ 
   \hline
East Asian & 1 & 32 & 872 & 67 &  &  \\ 
   \hline
East Asian & 1 & 33 & 1,328 & 155 & 57 & 23 \\ 
   \hline
East Asian & 1 & 34 & 1,865 & 298 & 82 & 37 \\ 
   \hline
East Asian & 1 & 35 & 2,059 & 688 & 111 & 57 \\ 
   \hline
East Asian & 1 & 36 & 1,111 & 540 & 62 & 46 \\ 
   \hline
East Asian & 2 & 32 & 259 & 10 &  &  \\ 
   \hline
East Asian & 2 & 33 & 462 & 20 &  &  \\ 
   \hline
East Asian & 2 & 34 & 741 & 92 &  &  \\ 
   \hline
East Asian & 2 & 35 & 1,053 & 273 & 62 & 33 \\ 
   \hline
East Asian & 2 & 36 & 679 & 242 &  &  \\ 
   \hline
East Asian & 3 & 32 & 169 & 3 &  &  \\ 
   \hline
East Asian & 3 & 33 & 279 & 16 &  &  \\ 
   \hline
East Asian & 3 & 34 & 468 & 47 &  &  \\ 
   \hline
East Asian & 3 & 35 & 638 & 118 &  &  \\ 
   \hline
East Asian & 3 & 36 & 480 & 182 &  &  \\ 
   \hline
East Asian & 4 & 32 & 103 & 3 &  &  \\ 
   \hline
East Asian & 4 & 33 & 179 & 11 &  &  \\ 
   \hline
East Asian & 4 & 34 & 326 & 35 &  &  \\ 
   \hline
East Asian & 4 & 35 & 434 & 90 &  &  \\ 
   \hline
East Asian & 4 & 36 & 348 & 109 &  &  \\ 
   \hline
East Asian & 5 & 32 & 66 & 2 &  &  \\ 
   \hline
East Asian & 5 & 33 & 80 & 7 &  &  \\ 
   \hline
East Asian & 5 & 34 & 131 & 13 &  &  \\ 
   \hline
East Asian & 5 & 35 & 186 & 41 &  &  \\ 
   \hline
East Asian & 5 & 36 & 156 & 62 &  &  \\ 
   \hline
East Asian & 6 & 32 & 175 & 3 &  &  \\ 
   \hline
East Asian & 6 & 33 & 248 & 17 &  &  \\ 
   \hline
East Asian & 6 & 34 & 391 & 43 &  &  \\ 
   \hline
East Asian & 6 & 35 & 562 & 99 &  &  \\ 
   \hline
East Asian & 6 & 36 & 435 & 153 &  &  \\ 
   \hline
East Asian & 7 & 32 & 109 & 3 &  &  \\ 
   \hline
East Asian & 7 & 33 & 200 & 11 &  &  \\ 
   \hline
East Asian & 7 & 34 & 375 & 39 &  &  \\ 
   \hline
East Asian & 7 & 35 & 651 & 108 &  &  \\ 
   \hline
East Asian & 7 & 36 & 508 & 158 &  &  \\ 
   \hline
East Asian & 8 & 32 & 110 & 4 &  &  \\ 
   \hline
East Asian & 8 & 33 & 164 & 12 &  &  \\ 
   \hline
East Asian & 8 & 34 & 272 & 31 &  &  \\ 
   \hline
East Asian & 8 & 35 & 299 & 77 &  &  \\ 
   \hline
East Asian & 8 & 36 & 181 & 73 &  &  \\ 
   \hline
East Asian & 9 & 32 & 1,196 & 37 & 53 & 3 \\ 
   \hline
East Asian & 9 & 33 & 2,044 & 101 & 92 & 19 \\ 
   \hline
East Asian & 9 & 34 & 3,302 & 262 & 163 & 45 \\ 
   \hline
East Asian & 9 & 35 & 4,130 & 599 & 198 & 65 \\ 
   \hline
East Asian & 9 & 36 & 2,277 & 521 & 100 & 43 \\ 
\hline
\caption{
    Aggregated counts of East Asian applicants and admits across groups defined by geography, equivalent ACT score, and legacy status.
    Admission is proxied by observing whether a final transcript is sent to one of the schools we consider.
    ``ZIP 1'' refers to the first digit of the student's high school ZIP code.
    To preserve confidentiality, legacy and non-legacy applicant cell counts with fewer than 50 applicants are redacted, along with the corresponding count of admits.
    Further, legacy and non-legacy admit cell counts of 0 are redacted, along with the corresponding count of applicants.
}   
\label{fig:replication_data_east_asian}
\end{longtable}

\FloatBarrier

\newpage

\restoregeometry

\begin{longtable}{|c|c||c|c||c|c|}
    \hline
 Ethnicity & ACT & Non-legacy applicants & Non-legacy admits & Legacy applicants & Legacy admits \\ 
 \hline
 White & 32 & 15,871 & 896 & 1,415 & 294 \\ 
   \hline
White & 33 & 22,291 & 1,735 & 2,459 & 591 \\ 
   \hline
White & 34 & 25,555 & 3,255 & 3,387 & 1,225 \\ 
   \hline
White & 35 & 21,762 & 4,354 & 3,480 & 1,630 \\ 
   \hline
White & 36 & 7,936 & 2,338 & 1,613 & 1,021 \\ 
   \hline
South Asian & 32 & 2,372 & 77 &  &  \\ 
   \hline
South Asian & 33 & 4,158 & 201 & 70 & 11 \\ 
   \hline
South Asian & 34 & 6,211 & 512 & 123 & 42 \\ 
   \hline
South Asian & 35 & 7,644 & 1,177 & 159 & 78 \\ 
   \hline
South Asian & 36 & 4,142 & 1,086 & 73 & 43 \\ 
   \hline
Southeast Asian & 32 & 1,303 & 54 &  &  \\ 
   \hline
Southeast Asian & 33 & 1,825 & 144 & 51 & 7 \\ 
   \hline
Southeast Asian & 34 & 2,029 & 203 & 75 & 15 \\ 
   \hline
Southeast Asian & 35 & 1,904 & 345 & 81 & 32 \\ 
   \hline
Southeast Asian & 36 & 701 & 198 &  &  \\ 
   \hline
East Asian & 32 & 3,557 & 173 & 168 & 22 \\ 
   \hline
East Asian & 33 & 5,836 & 433 & 268 & 69 \\ 
   \hline
East Asian & 34 & 9,337 & 1,070 & 467 & 146 \\ 
   \hline
East Asian & 35 & 12,225 & 2,695 & 599 & 263 \\ 
   \hline
East Asian & 36 & 7,737 & 2,752 & 301 & 185 \\ 
   \hline
\caption{
    Aggregated counts of legacy and non-legacy applicants and admits across ethnicity and ACT groups.
    Admission is proxied by observing whether a final transcript is sent to one of the schools we consider.
    To preserve confidentiality, legacy and non-legacy applicant cell counts with fewer than 50 applicants are redacted, along with the corresponding admit count.
    Further, legacy and non-legacy admit cell counts of 0 are redacted, along with the corresponding count of applicants.
} 
\label{fig:aggregated_replication_data}
\end{longtable}

\FloatBarrier

\newpage

\FloatBarrier

\newpage

\FloatBarrier

\newpage

\FloatBarrier

\newpage

\FloatBarrier

\newpage

\FloatBarrier

\newpage

\end{document}